\documentclass[aps,prl,twocolumn,showpacs,floatfix,preprintnumbers,amsmath,amssymb,superscriptaddress,reprint,epsfig]{revtex4-1}
\usepackage{epsfig}
\usepackage{graphicx}
\usepackage{amsmath}
\usepackage{physics}
\usepackage{rotating}
\usepackage{xcolor}
\usepackage[colorlinks,bookmarks=false,citecolor=blue,linkcolor=blue,urlcolor=blue]{hyperref}
\usepackage[all]{hypcap}   

\usepackage[final]{pdfpages}
\usepackage{pgffor}
\makeatletter
\AtBeginDocument{\let\LS@rot\@undefined}
\makeatother

\begin{document} 
\title{Interaction-induced directed transport in quantum chaotic systems}
\author{Sanku Paul}
\email{sankup005@gmail.com}
\affiliation{Department of Physics and Astronomy, Michigan State University, East Lansing, Michigan 48824, USA}
\author{J. Bharathi Kannan}
\email{jbharathi.kannan@students.iiserpune.ac.in}
\affiliation{Indian Institute of Science Education and Research, Dr. Homi Bhabha Road, Pune 411 008, India}
\author{M. S. Santhanam}
\email{santh@iiserpune.ac.in}
\affiliation{Indian Institute of Science Education and Research, Dr. Homi Bhabha Road, Pune 411 008, India}

\date{\today}

\begin{abstract}
Quantum directed transport can be realized in non-interacting, deterministic, chaotic systems by appropriately breaking the spatio-temporal symmetries in the potential. In this work, the focus is on the class of {\it interacting} quantum systems whose classical limit is chaotic. In this limit, one subsystem effectively acts as a source of ``noise'' to the other leading to temporal symmetry breaking.
Thus, the quantum directed currents can be generated with two ingredients -- broken spatial symmetry in the potential and presence of interactions.
This is demonstrated in two-body interacting kicked rotor and kicked Harper models. Unlike earlier schemes employed for single-particle ratchet currents, this work provides a minimal framework for realizing quantum directed transport in interacting systems. This can be generalized to many-body quantum chaotic systems. 
\end{abstract}

\maketitle

Basic physics teaches us that a particle can display directed motion only if it is acted upon by a net force $F \ne 0$. Ratchet effect is a counter-intuitive phenomenon that arises as an exception to this general rule \cite{Smo27, Fey66,Rei02,Hanggi_2009}. A diffusing particle, maintained away from equilibrium, can exhibit directed motion or ratchet effect even if the suitably averaged net force acting on it is zero, provided certain spatio-temporal symmetries are broken. This amounts to extracting useful work out of thermal fluctuations in {\it non-equilibrium} systems and is not prohibited by the second law of thermodynamics. Apart from clarifying the foundational principles of thermodynamics \cite{Fey66}, the ratchet mechanisms are the drivers of many biological processes, for instance, the intracelluar transport of molecules such as kinesin along microtubules, muscular contractions and movement of bacteria in a suspension fluid \cite{Ast2007a}. These natural processes have inspired a plethora of ratchet models and experiments \cite{JulAjdPro97,Bermudez2000,Rei02,Villegas2003,Hanggi_2005,SjoPetDio2006,Serreli2007,Mahmud2009,CubRen2016,ArzVilVol2017,MukXieLie2018,For2021} in all the areas of physical sciences. This includes a general class of dissipative ratchets \cite{Rei02,Carlo_2005,Hanggi_2009}, electron ratchets \cite{Linke_1999,LauKed2020}, and recent applications for enhancing the efficiency of photovoltaic cells \cite{YosEkiFar2012,PusYosHyl2016,SogHunTam2021}.

These developments have largely focussed on exploiting thermal noise for extracting useful work. In principle, {\it clean} and {\it noise-free} ratchets can be created using Hamiltonian systems if they display chaotic dynamics. In these models, the inherent classical chaos formally plays the role of thermal noise within the framework of deterministic dynamics. This has led to explorations of classical and quantum ratchets using chaotic Hamiltonian models such as the kicked rotor. 
In general, two distinct types of Hamiltonian models can generate directed currents. One approach requires bounded  classical phase space and a co-existence of regular and chaotic dynamics \cite{DenMorFla07, FlaYevZol00, DenFlaHan14, SchOttKet01}. These are subject to a semi-classical sum rule and directed momentum current arises due to the net difference between the currents carried by regular and chaotic phase space regions. In the $\hbar \to 0$ limit, this carries over to the quantum regime as well. The second approach accommodates nearly complete chaotic phase space and requires manipulating the kick sequences to break the temporal symmetry \cite{MonDanHut02,JonGooMea07}. This mechanism supports quantum ratchet currents if dynamical localization occurs, the quantum kicked rotor being a prime example. Another variant is the ratchet accelerators in which directed currents increase linearly with time under conditions of quantum resonance \cite{LunWal05,PolCarGab07,Dana_2008}.

In the last two decades, the classical and quantum dynamics of single particle chaotic Hamiltonian ratchets were extensively studied \cite{GonBru04, LunWal05, PolCarGab07, SchOttKet01, DenMorFla07}.  
In contrast, despite the exploding interest in interacting quantum many-body systems, their ratchet dynamics largely remains unexplored. Theoretical proposals based on periodically kicking the condensates indicate that the mean-field interactions can induce directed current \cite{PolBenCas2007,ZhaDinLiu2016}. Quantum directed current was also experimentally observed in a driven Bose-Einstein condensate in a toroidal trap when spatio-temporal symmetries were broken \cite{SalKliHec2009}. Recently, it was shown that in the ratchet regime quantum-classical correspondence differs from the generically expected delocalization of states and maximal entanglement applicable for a many-body quantum chaotic system \cite{ValShcHel2018,ValShcSol2019}. In all these examples, the dynamics of condensates in the mean-field limit is governed by a nonlinear Schr\"odinger equation. 
Ironically, within the scope of {\it linear} quantum dynamics, not much is known about the interplay between the interactions and classical chaos in generating quantum ratchet currents.

Hence, we consider a simple prototype of two independent particles represented as chaotic systems coupled by an interaction potential, whose quantum dynamics is described by the linear Schr\"odinger equation.  In the chaotic limit, any one of the particles can play the role of ``environment'' to the other \cite{SanArn20}. Thus, an interacting system effectively includes its own ``environment'', implying that breaking only the spatial symmetries can induce ratchet currents. This is because, as we analytically show in this paper, the temporal symmetries in the subsystem are inherently broken by the inter-particle interactions.
Remarkably, the interacting models provide a novel paradigm for generating directed currents in quantum systems whose classical counterparts display complete chaos. It discards the requirement of mixed classical phase space or manipulation of kick sequences or parameters set for quantum resonances. To emphasize its generic nature, we demonstrate ratchet currents in two distinct interacting potentials, one which obeys the assumption of the Kolmogorov-Arnold-Moser (KAM) theorem \cite{LicLie2013} while the other violates an assumption of KAM theorem \cite{Gardiner_1997,Sankarnarayanan_2001,Sanku_2016,Sanku_2018}. 

We thus propose interacting models as natural candidates for realizing directed quantum currents.
A general two-particle Hamiltonian of the form
\begin{equation}
H  = H_1(q_1,p_1,t) + H_2(q_2,p_2,t) + \varepsilon ~ V_{\rm int}(q_1,q_2) ~ f(t)
\label{ham1}
\end{equation}
is considered, where $H_i(q_i,p_i,t)$ with $i=1,2$ represents a single-particle periodically kicked chaotic sub-system labeled 1 and 2, and $f(t) = \sum_{n=-\infty}^{\infty} \delta(t-n)$. 
The interaction potential $V_{\rm int}(q_1,q_2)$ of strength $\varepsilon$ is such that if $\varepsilon=0$ the system reduces to two independent chaotic systems. Quantum dynamics can be conveniently generated by the unitary period-1 time-evolution operator $\mathcal{U} = (U_1 \otimes U_2) U_{\rm int}$, where $U_i = e^{-iH_i/\hbar_s}$  
with $i=1,2$ represents the evolution operator for particle 1 and 2 respectively, while $U_{\rm int}=e^{-i \varepsilon \frac{V_{\rm int}}{\hbar_s}}$ arises from the interaction and $\hbar_s$ is the scaled Planck constant which is set to $1$ throughout this paper. In this paper, we show that if $V_{\rm int}(q_1,q_2)$ is chosen to break the spatial symmetry, then directed current emerges as a result of interactions without requiring an explicit temporal symmetry breaking. 

This is demonstrated in two chaotic systems -- coupled kicked rotor (CKR) and kicked Harper (CKH) models defined in a cylindrical phase space. Their Hamiltonians, respectively, are given by
\begin{align}
H_i^{\rm KR}(q_i,p_i,t) & = p_i^2 / 2 + K_i^r \cos q_i \sum_n \delta(t-n)\,,  \mbox{~ and} \label{ham2} \\
H_i^{\rm KH}(q_i,p_i,t) & = L_i^h\cos p_i + K_i^h \cos q_i \sum_n \delta(t-n)\,,
\label{ham3}
\end{align}
with $i=1,2$.
Kicked rotor is a well-studied model of Hamiltonian chaos \cite{Casati_1979,IZRAILEV1990,SanPauKan2022} and corresponds to a particle receiving periodic kicks. In the quantum regime, a generic feature is the emergence of dynamical localization that suppresses classical diffusive dynamics for kick strengths $K_r \gg 1$. This is analogous to the Anderson localization observed in disordered lattices \cite{FisShmGre82}. The kicked Harper model is a chaotic system and is physically related to the electronic motion in 2D crystal with an external magnetic field \cite{Sokoloff_1985,Lima_1991}. Its quantum version can display localization or delocalization effects depending on the choice of $K^h$ and $L^h$\cite{Lima_1991}. To avoid any permutation symmetries in the system, we choose $K_1^{r,h}\neq K_2^{r,h}$ and $L_1^h\neq L_2^h$. In this work, we fix the kick strengths of CKR as $K_1^r=1.5$, $K_2^r=0.8$, and the parameters for the CKH model are $L_1^h=4$, $L_2^h=4.2$, $K_1^h=2$, and $K_2^h=2.1$, to remain in chaotic regime.

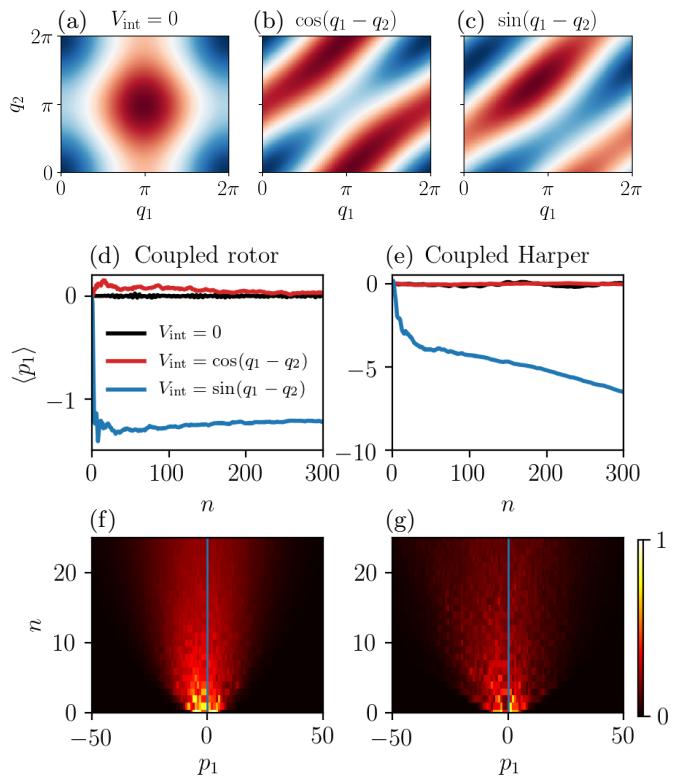
\begin{figure}[t]
    \centering
    \includegraphics[width=\linewidth]{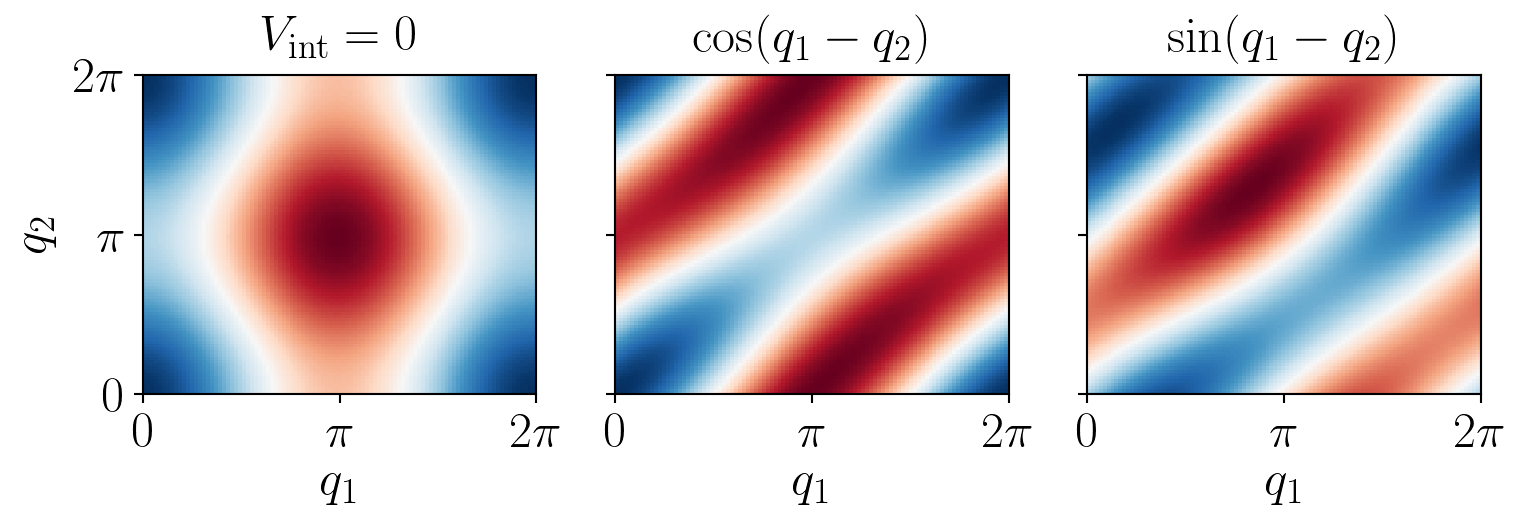}
    \includegraphics[width=1.05\linewidth]{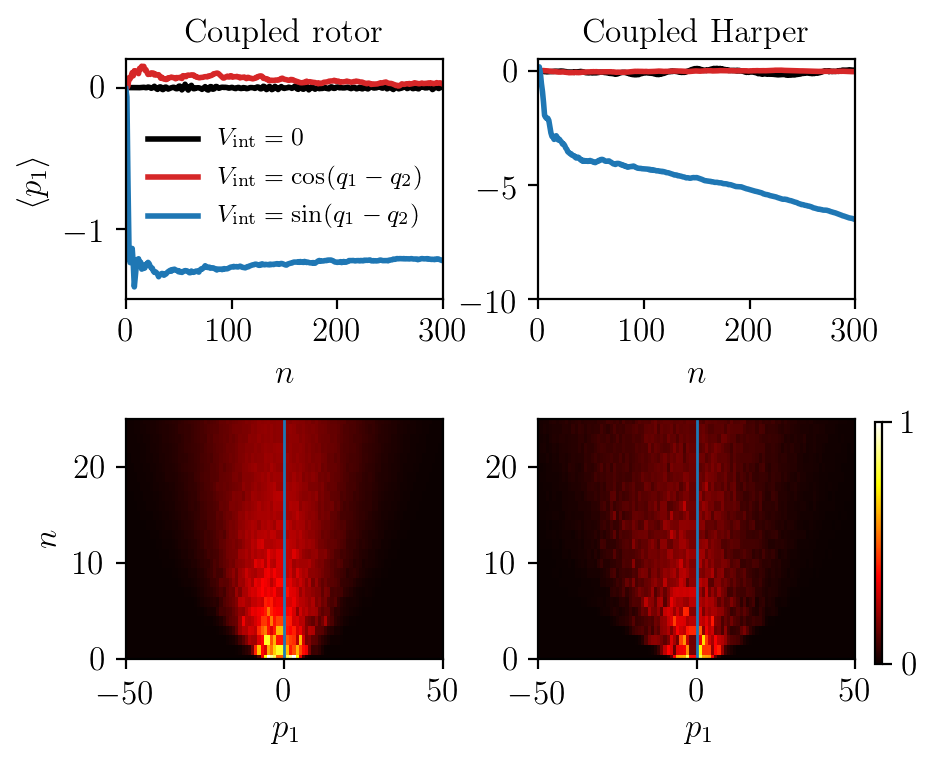}
	\begin{picture}(0,0)
	\put(-100,300){(a)}
	\put(-26,300){(b)}
	\put(51,300){(c)}
	\put(-88,211){(d)}
	\put(24,211){(e)}
	\put(-88,111){(f)}
	\put(24,111){(g)}
	\end{picture}	
	\vspace{-0.3cm}	
    \caption{(a-c) Color maps of the effective potential $V_{\rm eff}=K_1^r \cos q_1 + K_2^r \cos q_2 + \varepsilon V_{\rm int}(q_1,q_2)$ for three cases with $\varepsilon=5$ ; (a) $V_{\rm int}=0$, (b) $\ V_{\rm int}=\cos(q_1-q_2)$, and (c) $V_{\rm int}=\sin(q_1-q_2)$. Note that (c) lacks spatial symmetry.
    (d) Mean current $\langle p_1 \rangle$ for the same three cases is shown for $V_{\rm int}=0$ (black line), $V_{\rm int}=\cos(q_1-q_2)$ (red), and $V_{\rm int}=\sin(q_1-q_2)$ (blue).  
    Average is taken over $200$ different quasi-momenta $\beta$ taken randomly from an interval $[-0.1,0.1]$. Momentum distribution $f(p_1)$ scaled by its peak value is shown for (f) coupled kicked rotor, and (g) coupled Harper model. Vertical blue line at $p_1=0$ is a guide to the eye. Left panel corresponds to coupled kicked rotor and right panel to coupled Harper model. Parameters for coupled kicked rotor are $K_1^r=1.5$, and $K_2^r=0.8$ while that for coupled Harper are $K_1^h=2, K_2^h=2.1, L_1^h=4,$ and $L_2^h=4.2$.
    }
    \label{fig:symm1}
\end{figure}

For both systems, the initial state is chosen to be a direct-product state $|\psi(0)\rangle = |\phi_1(0) \rangle \otimes |\phi_2(0)\rangle$, where $|\phi_i(0)\rangle = 1/\sqrt{2\pi}, (i=1,2)$ being the zero momentum state. The time evolved state is $|\psi(n)\rangle = \mathcal{U}^n |\psi(0)\rangle$. The evolution operator takes the form $U_i=U_i^{\rm free}(p_i+\beta_i)U_i^{\rm kick}$,
where $\beta_i$ is the corresponding quasi-momentum for particle labeled $i=1,2$. In this, $U_i^{\rm free}$ is the free evolution operator while $U_i^{\rm kick}$ corresponds to the kicking part. We focus on the mean quantum current generated in subsystem-1 : $\left\langle p_1 \right\rangle=\left\langle\psi(n)\left| p_1 \right| \psi(n)\right\rangle$. This is numerically computed and is averaged over several quasi-momenta.

Figure \ref{fig:symm1} (a,b,c), displays the color map of the effective potential, namely, $V_{\rm eff} = K^s_1 \cos q_1 + K^s_2 \cos q_2 + \varepsilon V_{\rm int}(q_1,q_2)$  of Eq. \ref{ham1}, where $s=h$ or $r$. 
For the given kicking potentials in Eqs. \ref{ham2}-\ref{ham3}, three different cases are considered : ({\sl i}) no interactions, $\varepsilon=0$, and spatial symmetry is maintained (Fig. \ref{fig:symm1}(a)), ({\sl ii}) cosine interaction, $\varepsilon \ne 0$ and $V_{\rm int}(q_1,q_2) = \cos(q_1-q_2)$, spatial symmetry survives  (Fig. \ref{fig:symm1}(b)), and ({\sl iii}) sine interaction, $\varepsilon \ne 0$ and $V_{\rm int}(q_1,q_2) = \sin(q_1-q_2)$, spatial symmetry is broken (Fig. \ref{fig:symm1}(c)). For these three cases, Fig. \ref{fig:symm1}(d,e) displays $\left\langle p_1 \right\rangle$ for the coupled kicked rotor and coupled Harper systems. Notice that, in both systems, directed current is realized, {\it i.e.}, $\left\langle p_1 \right\rangle \ne 0$ only in the case of sine interactions with broken spatial symmetry. Surprisingly, this did not require explicit temporal symmetry breaking. Notice that the directed current is absent for the cosine interaction since the spatial symmetry is preserved. Furthermore, directed current is also absent if $V_{\rm int} = 0$ (blue curve) indicating that spatial-symmetry breaking interactions are crucial for generating directed currents.

In Fig. \ref{fig:symm1}(f,g), the time evolution of momentum distributions $f(p_1)$ with  $V_{\rm int}(q_1,q_2) = \sin(q_1-q_2)$, are shown for CKR and CKH systems. It reveals that the momentum distributions are asymmetric about $p_1=0$ or for that matter about any $p_1$. In CKR, the asymmetry freezes quickly leading to a saturated directed current as seen in fig. \ref{fig:symm1}(d). In contrast, in the CKH model, the asymmetry continues to grow and leads to a directed current with acceleration as shown in fig. \ref{fig:symm1}(e). This is due to the ballistic energy growth (not shown here) observed for the parameters used in this work \cite{Lima_1991}. The sign of the generated current -- negative in these cases in Fig. \ref{fig:symm1}(d,e) -- depends on the choice of parameters.

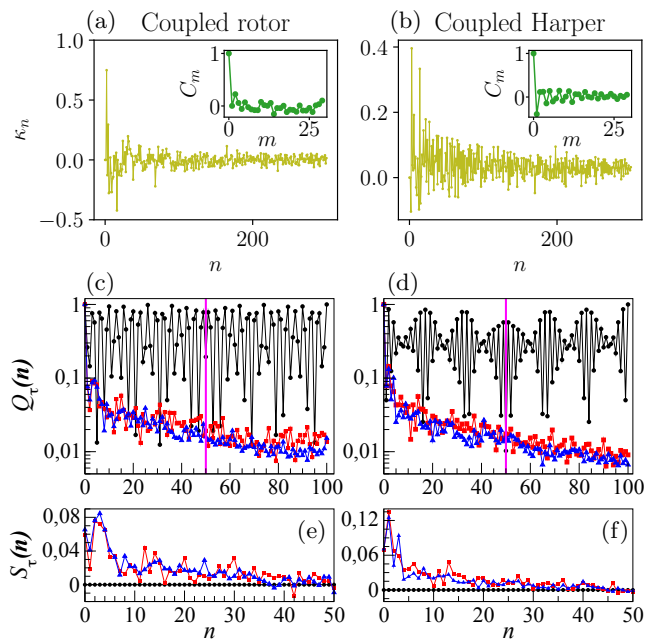
\begin{figure}[t]
    \centering
    \includegraphics[width=\linewidth]{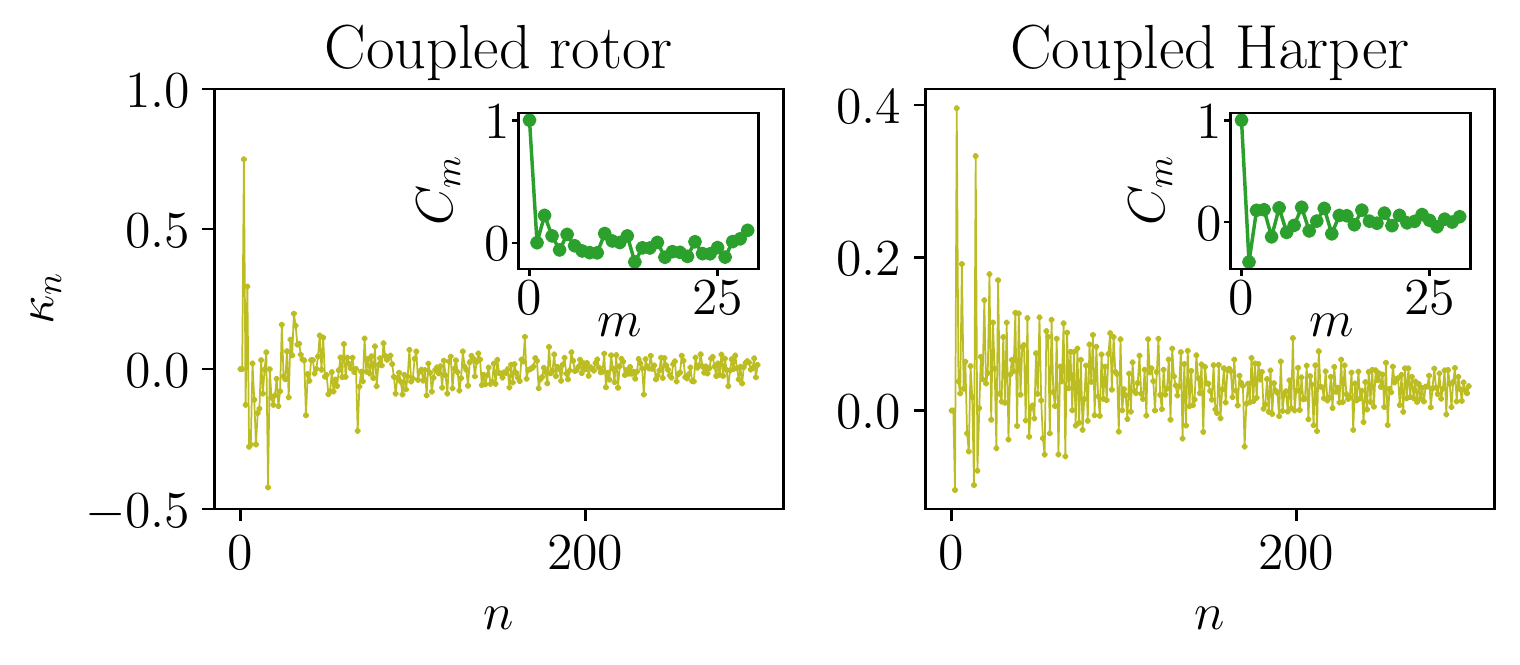}
    \includegraphics[width=\linewidth]{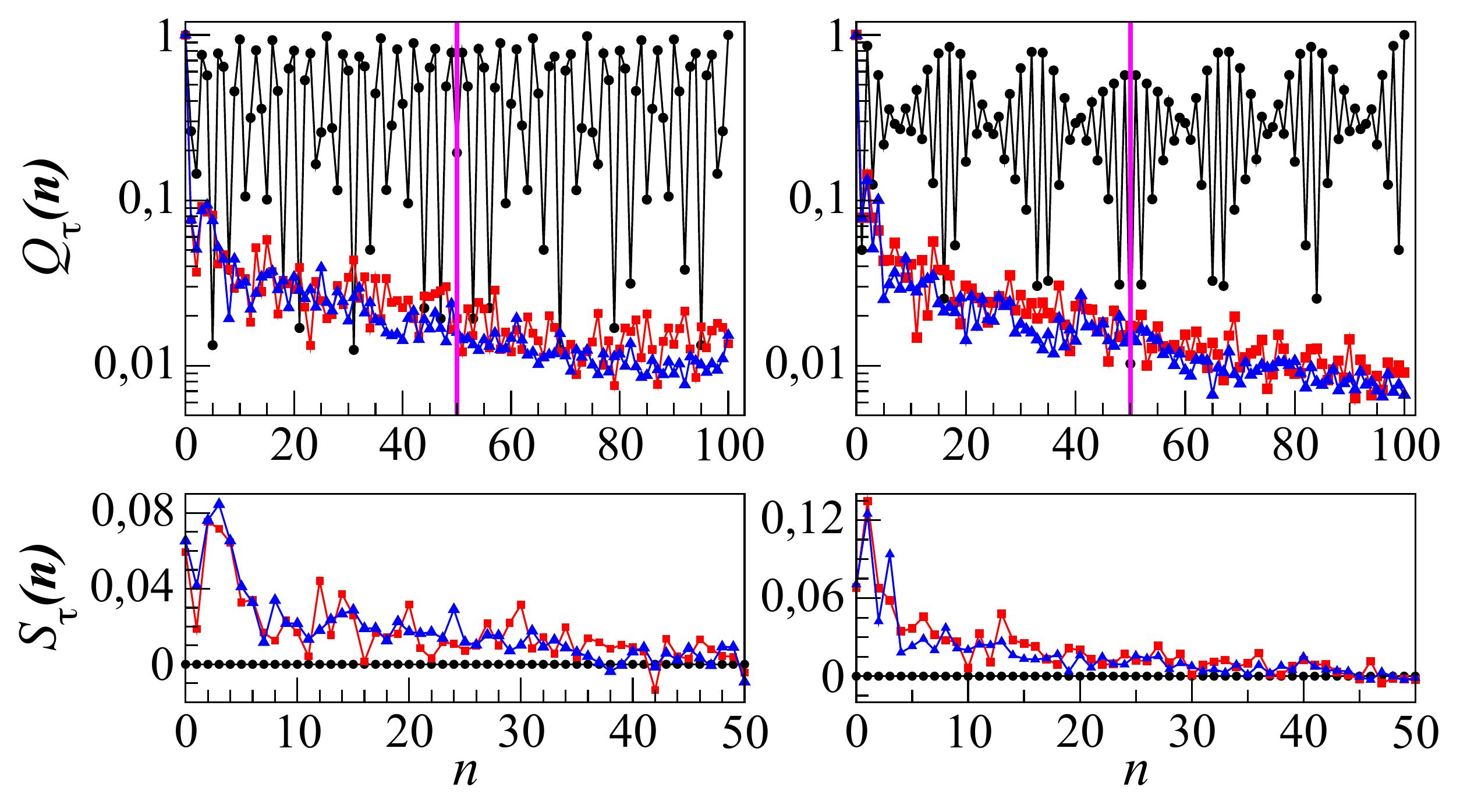}
	\begin{picture}(0,0)
	\put(-90,246){(a)}
	\put(25,246){(b)}
	\put(-90,147){(c)}
	\put(24,147){(d)}
	\put(-10,53){(e)}
	\put(105,53){(f)}
	\end{picture}
	\vspace{-0.2cm}	
    \caption{(a,b) Effective kick strength $\kappa_n = \langle \psi(n) | \varepsilon \sin q_j | \psi(n) \rangle$ as a function of time. The insets show the numerically computed autocorrelation function $C_m = \langle \kappa_n \kappa_{n+m} \rangle$. 
    (c,d) $Q_{\tau}(n)$ with no interaction $\varepsilon=0$ (black symbols), with interactions $V_{\rm int}=\cos(q_1-q_2)$ (red), and $V_{\rm int}=\sin(q_1-q_2)$ (blue). The vertical magenta line indicates $n=\tau$ about which time reversal symmetry is expected. (e,f) $S_{\tau}(n)$ indicating the existence or absence of time-reversal symmetry about $n=\tau$. Note that temporal symmetry is preserved if $\varepsilon=0$, and it is broken for $\varepsilon \ne 0$. Other parameters are the same as in Fig. \ref{fig:symm1}$\,.$
    }
    \label{fig:qcurr1}
\end{figure}

From these discussions, it is clear that the breaking of spatial symmetry alone can produce directed currents in an interacting system. Then, a natural question arises about the enabling role of chaos in generating the currents. In fact, chaos is a necessary ingredient for making one subsystem effectively act as an noisy environment for the other. This, in turn, breaks the temporal symmetry of the individual subsystems. To argue that one subsystem is a source of noisy environment to the other, we expand $V_{\rm int}=\sin(q_1-q_2)$ and rearrange the terms to identify an effective kicking strength for the individual subsystem as $K_i^{\rm eff}=K_i+(1-2\Theta(i-j))\varepsilon \sin q_j$ where $i\neq j$ and $\Theta(.)$ is a Heaviside step function. The numerical simulations in Fig. \ref{fig:qcurr1}(a,b), show that if the system is chaotic, then $\kappa_n = \langle \psi(n) | \varepsilon \sin q_j | \psi(n) \rangle$ has the characteristics of a ``noisy'' process. In particular, its autocorrelation function is $C_m=\langle \kappa_n \kappa_{n+m} \rangle \sim A ~\delta(m)$ ($A$ is a constant), representing an uncorrelated process. Thus, $K^{\rm eff}_i$ experienced by a subsystem is driven by the noisy dynamics of the other subsystem implying that one subsystem effectively acts as an ``environment'' for the other.

Next, we analytically show that the interaction breaks the temporal symmetry of the individual subsystems. Let us consider an initial state at time $n=0$ to be a direct product state of the form
$|\psi(0)\rangle = |\phi_1(0)\rangle \otimes |\phi_2(0)\rangle$, where $|\phi_i(0)\rangle, (i=1,2)$ are the initial states in subsystems $H_1$ and $H_2$ respectively. To demonstrate that the interactions induce breaking of time-reversal symmetry, the following sequence of operations are performed : the initial state is evolved forward for duration $\tau > 0$ under the action of $H$ in Eq. \ref{ham1}. Next, particle-1 is evolved backward in time for duration $-\tau$, while particle-2 is evolved forward for duration $\tau$. This sequence of operations can be represented as
\begin{equation}
 |\psi(\tau')\rangle = [(U_1^{-\tau} \otimes U_2^{\tau}) U_{\rm int}^\tau] ~ [(U_1 \otimes U_2) U_{\rm int} ]^\tau ~ |\psi(0)\rangle\,.
\end{equation}
In this, $\tau'$ indicates successive time duration of $(\tau,-\tau)$ for particle-1, and $(\tau,\tau)$ for particle-2.
Let $\rho(\tau') =  |\psi(\tau')\rangle \langle \psi(\tau')|$ be the associated density matrix, and the reduced density matrix for particle-1 is $\rho_1(\tau') = \mbox{Tr}_2~\rho(\tau')$. Due to unitary evolution, we might expect particle-1 to retrace its path to the initial state.
As shown exactly in the supplementary information, this expectation is true in the absence of interactions ($\varepsilon=0$). Using the definition $Q_{\tau}(n)=\langle \phi_1(0) | \rho_1(n) | \phi_1(0) \rangle$, we have that $Q_{\tau'} = 1$. 
In this case, existence of temporal symmetry about $n=\tau$ implies precise time-reversal, {\it i.e.}, $Q_{\tau}(n) = Q_{\tau}(2\tau-n), n=1,2 \dots \tau$.
If the interaction is present for $\varepsilon \ne 0$, it can be exactly shown (see supplementary material) that
\begin{equation}
0 < Q_{\tau}(n=\tau') = \langle \phi_1(0) | \rho_1(\tau') | \phi_1(0) \rangle < 1\,.
\end{equation}
Indeed, $Q_{\tau}(n) \ne Q_{\tau}(2\tau-n)$.
These results confirm that interactions induce the breaking of time-reversal symmetry. This is a general feature of interacting systems and not specific to kicked models alone.

Based on the computed $Q_{\tau}(n)$ shown in Fig. \ref{fig:qcurr1}(a,b), it is clear that the temporal symmetry is absent if $\varepsilon \ne 0$. The simulations in Fig. \ref{fig:qcurr1}(c,d) with $\tau=50$ show $Q_{50}(n)$ for $n=1,2\dots 2\tau$ iterations. With $\varepsilon=0$ (black circles), under time-reversal, the state of subsystem-1 retraces its path to the initial state. In particular, for both CKR and CKH, notice that the time-evolution is symmetric about $\tau=50$ (marked as magenta line). In contrast, if $\varepsilon \ne 0$, $Q_{50}(100) < 1$ (red and blue symbols in Fig. \ref{fig:symm1}(d-e), corresponding to cosine and sine interaction terms) and a lack of temporal symmetry is clearly evident. Figure \ref{fig:qcurr1}(e,f), for CKR and CKH, shows a diagonostic measure $S_{\tau}(n) = Q_{\tau}(n) - Q_{\tau}(2\tau-n), n=1,2 \dots \tau$ to reveal the existence or absence of temporal symmetry. Clearly, $S_{\tau}(n)=0$ if temporal symmetry about $n=\tau$ is present, while $S_{\tau}(n) \ne 0$ if this symmetry is absent. As seen in Fig. \ref{fig:qcurr1}(e,f), $S_{\tau}(n)=0$ for $\varepsilon=0$, and $S_{\tau}(n) \ne 0$ for $\varepsilon \ne 0$.
With appropriate choice of symmetry-broken interaction $V_{\rm int}(q_1,q_2)$, generically, directed current will be generated. This is confirmed by the simulations shown in Fig. \ref{fig:symm1}(d,e).

\begin{figure}[t]
    \centering
    \includegraphics[width=\linewidth]{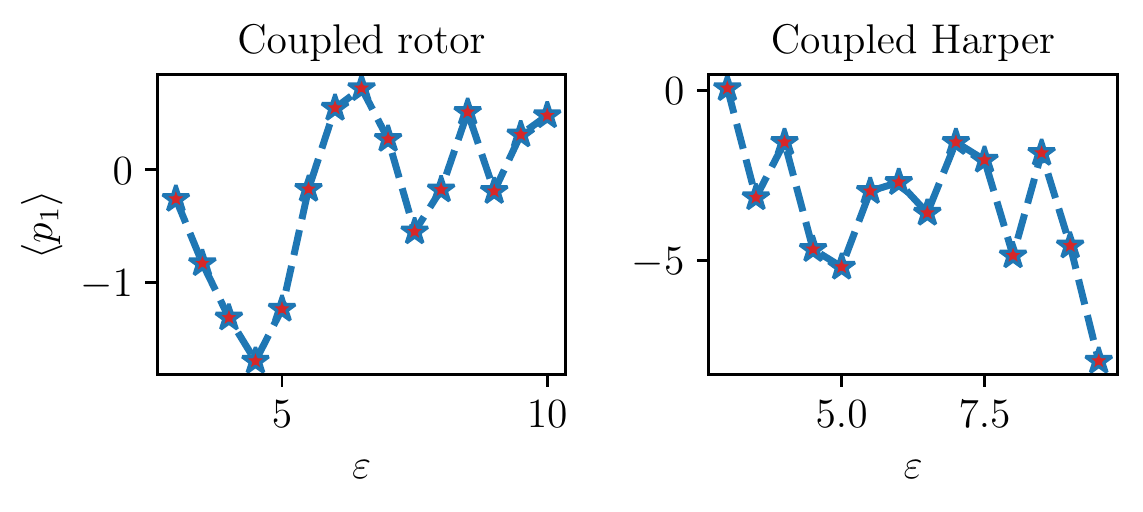}
	\begin{picture}(0,0)
	\put(-90,109){(a)}
	\put(28,109){(b)}
	\end{picture}	
	\vspace{-0.6cm}
    \caption{Quantum current $\langle p_1 \rangle$ generated at fixed time $n=200$ by the coupled kicked rotor and kicked Harper system as a function of coupling strength $\varepsilon$. Other parameters are the same as in Fig. \ref{fig:symm1}$\,.$}
    \label{fig:qcurr_vs_K}
\end{figure}
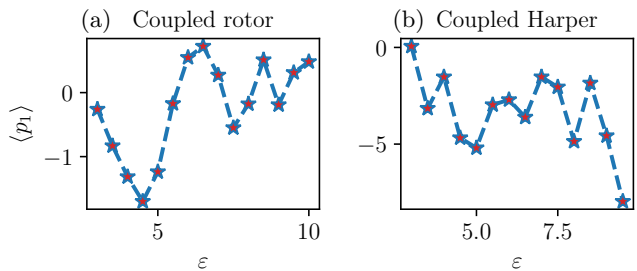

To obtain a global perspective, Fig. \ref{fig:qcurr_vs_K} shows mean current after $n=200$ iterations as a function of coupling strength $\varepsilon$ for both the models. In both cases, interaction potential $V_{\rm int}=\sin(q_1-q_2)$ is so chosen to break the spatial symmetry. It is clear that for any $\varepsilon > 0$ directed currents are generated in both the models. The interaction-induced currents are robust though they do not reveal any obvious systematic relation with $\varepsilon$.

In contrast to the interacting spin models, the system in Eq. \ref{ham1} has the advantage that it has an unambiguous classical limit that allows for a comparison with quantum regime. The classical dynamics of Eq. \ref{ham1} can be reduced to a stroboscopic map on a cylinder (see supplementary material). For sufficiently large kick strengths, the classical dynamics of kicked rotor and Harper model is strongly chaotic \cite{SanPauKan2022,Lima_1991}. Further, chaos dominates even for small kick strengths if $\varepsilon >> 1$ \cite{Arul_2001}. In this regime, the classical mean energy of either subsystem displays diffusive growth, {\it i.e.}, $\langle E \rangle \sim D_{\rm cl} n$, where $D_{\rm cl}$ is the classical diffusion coefficient. Though interactions induce directed current by breaking temporal symmetries in the {\it quantum} regime (see Fig. \ref{fig:symm1}), in the corresponding classical regime the nearly complete chaos and absence of temporal symmetry breaking precludes any possibility of directed currents. Thus, directed classical current is absent (see supplementary material) implying that the observed currents in Fig. \ref{fig:symm1} have a quantum origin. Thus, the interactions can be chosen to provide a minimal framework -- requiring only broken spatial symmetry -- to realize quantum ratchet currents in coupled systems. Taken together, these results lead to the following proposition: two ingredients are necessary to realize quantum directed currents, namely, the system $H$ in Eq. \ref{ham1} must be chaotic and the effective interaction potential must break spatial symmetry.

\begin{figure}[t]
    \centering
    \includegraphics[width=\linewidth]{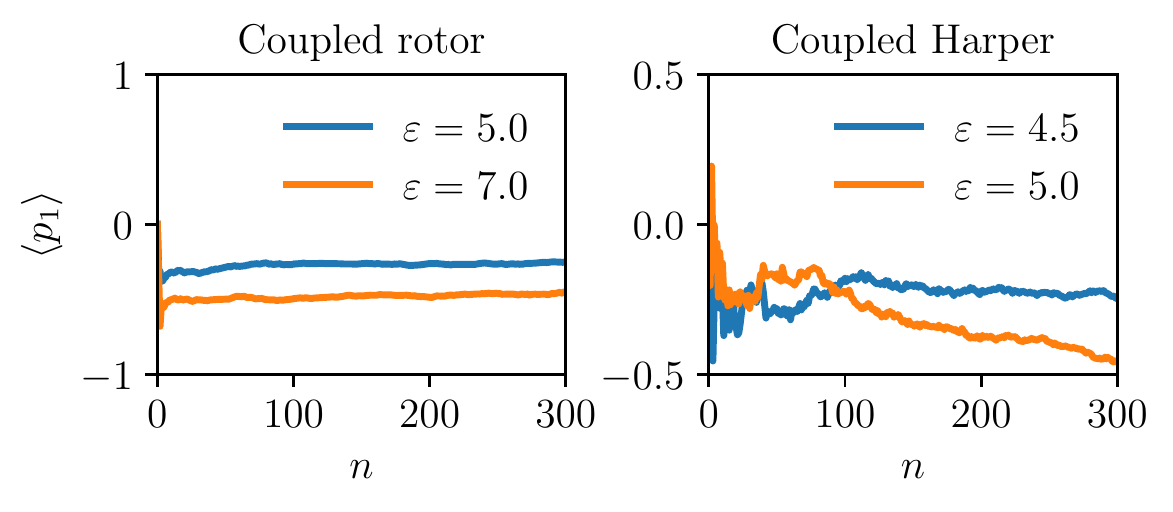}
	\begin{picture}(0,0)
	\put(-90,107){(a)}
	\put(24,107){(b)}
	\end{picture}	
	\vspace{-0.6cm}
    \caption{Mean quantum current $\langle p_1 \rangle$ shown as a function of time for the interaction
    potential $V_{\rm int}= - 2 \varepsilon |q_1| |q_2|$ in the system represented by Eq. \ref{ham1}$\,.$  Other parameters are the same as in Fig. \ref{fig:symm1}$\,.$}
    \label{fig:modint}
\end{figure}
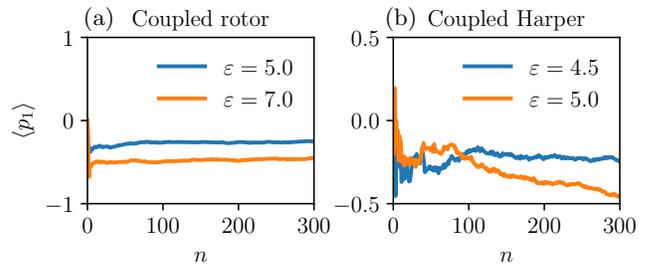

Finally, it must be emphasized that the results presented here are general and not specific to the choice of effective potential $V_{\rm eff}$, the only requirement being that $V_{\rm eff}$ must break the spatial symmetry. To underscore this point, Fig. \ref{fig:modint} displays quantum currents obtained for $V_{\rm eff}(q_1,q_2) = K_1^s \sin q_1 + K_2^s \sin q_2 + V_{\rm int}$, where $V_{\rm int} = - 2 \varepsilon |q_1| |q_2|$ and $s=r$ or $h$. This interaction potential is non-analytic and violates one of the assumptions of the KAM theorem and hence it is a non-KAM system \cite{Sankarnarayanan_2001,Sanku_2016,Sanku_2018}. The classical limit of this system is chaotic, and the interaction breaks the spatial symmetry of the potential. As shown in Fig. \ref{fig:modint}, quantum directed currents are generated in this case as well. Throughout this work, we have demonstrated quantum currents generated by subsystem-1, though similar results would be obtained had subsystem-2 been chosen. It might be feasible to realize the coupled kicked rotors in experiments by subjecting atomic matter waves to flashing incommensurate optical lattices, the interactions arise during the free evolution when the lattices are off \cite{Gadway_2013}. 

{\it Summary} : Directed currents can be usually generated if the relevant spatio-temporal symmetries in the potential are broken. We have shown analytically that if two quantum systems with a chaotic classical limit are coupled together, then the time-reversal symmetry breaking is induced by the interaction potential acting between the two subsystems. Effectively, one part of the system acts as the ``environment'' to the other part. Directed quantum currents can be generated by explicitly breaking only the spatial symmetry of the interaction potential. This has been demonstrated in two different coupled chaotic systems -- namely, the coupled kicked rotor and the coupled Harper model -- in a regime of pre-dominant classical chaos. Thus, chaos in the subsystems and presence of interactions among them leads to directed quantum currents, even if the classical currents average to zero. This interaction induced mechanism is sufficiently general and would also be applicable to other many-body interacting quantum systems even if they may not have a definite classical limit.

\acknowledgments
MSS would like to acknowledge the MATRICS grant from SERB, Govt of India. S.P. is supported by the start-up funding from Michigan State University.

\bibliographystyle{apsrev4-1}
\bibliography{ref}

\begin{thebibliography}{51}%
\makeatletter
\providecommand \@ifxundefined [1]{%
 \@ifx{#1\undefined}
}%
\providecommand \@ifnum [1]{%
 \ifnum #1\expandafter \@firstoftwo
 \else \expandafter \@secondoftwo
 \fi
}%
\providecommand \@ifx [1]{%
 \ifx #1\expandafter \@firstoftwo
 \else \expandafter \@secondoftwo
 \fi
}%
\providecommand \natexlab [1]{#1}%
\providecommand \enquote  [1]{``#1''}%
\providecommand \bibnamefont  [1]{#1}%
\providecommand \bibfnamefont [1]{#1}%
\providecommand \citenamefont [1]{#1}%
\providecommand \href@noop [0]{\@secondoftwo}%
\providecommand \href [0]{\begingroup \@sanitize@url \@href}%
\providecommand \@href[1]{\@@startlink{#1}\@@href}%
\providecommand \@@href[1]{\endgroup#1\@@endlink}%
\providecommand \@sanitize@url [0]{\catcode `\\12\catcode `\$12\catcode
  `\&12\catcode `\#12\catcode `\^12\catcode `\_12\catcode `\%12\relax}%
\providecommand \@@startlink[1]{}%
\providecommand \@@endlink[0]{}%
\providecommand \url  [0]{\begingroup\@sanitize@url \@url }%
\providecommand \@url [1]{\endgroup\@href {#1}{\urlprefix }}%
\providecommand \urlprefix  [0]{URL }%
\providecommand \Eprint [0]{\href }%
\providecommand \doibase [0]{http://dx.doi.org/}%
\providecommand \selectlanguage [0]{\@gobble}%
\providecommand \bibinfo  [0]{\@secondoftwo}%
\providecommand \bibfield  [0]{\@secondoftwo}%
\providecommand \translation [1]{[#1]}%
\providecommand \BibitemOpen [0]{}%
\providecommand \bibitemStop [0]{}%
\providecommand \bibitemNoStop [0]{.\EOS\space}%
\providecommand \EOS [0]{\spacefactor3000\relax}%
\providecommand \BibitemShut  [1]{\csname bibitem#1\endcsname}%
\let\auto@bib@innerbib\@empty
\bibitem [{\citenamefont {Smoluchowski}(1927)}]{Smo27}%
  \BibitemOpen
  \bibfield  {author} {\bibinfo {author} {\bibfnamefont {M.}~\bibnamefont
  {Smoluchowski}},\ }\href {http://eudml.org/doc/215787} {\bibfield  {journal}
  {\bibinfo  {journal} {Pisma Mariana Smoluchowskiego}\ }\textbf {\bibinfo
  {volume} {2}},\ \bibinfo {pages} {226} (\bibinfo {year} {1927})}\BibitemShut
  {NoStop}%
\bibitem [{\citenamefont {Feynman}\ \emph {et~al.}(1966)\citenamefont
  {Feynman}, \citenamefont {Leighton},\ and\ \citenamefont {Sands}}]{Fey66}%
  \BibitemOpen
  \bibfield  {author} {\bibinfo {author} {\bibfnamefont {R.~P.}\ \bibnamefont
  {Feynman}}, \bibinfo {author} {\bibfnamefont {R.~B.}\ \bibnamefont
  {Leighton}}, \ and\ \bibinfo {author} {\bibfnamefont {M.}~\bibnamefont
  {Sands}},\ }\href@noop {} {\emph {\bibinfo {title} {Feynman Lectures on
  Physics, Vol. 1,}}}\ (\bibinfo {year} {1966})\BibitemShut {NoStop}%
\bibitem [{\citenamefont {Reimann}(2002)}]{Rei02}%
  \BibitemOpen
  \bibfield  {author} {\bibinfo {author} {\bibfnamefont {P.}~\bibnamefont
  {Reimann}},\ }\href {\doibase https://doi.org/10.1016/S0370-1573(01)00081-3}
  {\bibfield  {journal} {\bibinfo  {journal} {Phys. Rep}\ }\textbf {\bibinfo
  {volume} {361}},\ \bibinfo {pages} {57} (\bibinfo {year} {2002})}\BibitemShut
  {NoStop}%
\bibitem [{\citenamefont {H\"anggi}\ and\ \citenamefont
  {Marchesoni}(2009)}]{Hanggi_2009}%
  \BibitemOpen
  \bibfield  {author} {\bibinfo {author} {\bibfnamefont {P.}~\bibnamefont
  {H\"anggi}}\ and\ \bibinfo {author} {\bibfnamefont {F.}~\bibnamefont
  {Marchesoni}},\ }\href {\doibase 10.1103/RevModPhys.81.387} {\bibfield
  {journal} {\bibinfo  {journal} {Rev. Mod. Phys.}\ }\textbf {\bibinfo {volume}
  {81}},\ \bibinfo {pages} {387} (\bibinfo {year} {2009})}\BibitemShut
  {NoStop}%
\bibitem [{\citenamefont {Astumian}(2007)}]{Ast2007a}%
  \BibitemOpen
  \bibfield  {author} {\bibinfo {author} {\bibfnamefont {R.~D.}\ \bibnamefont
  {Astumian}},\ }\href {\doibase 10.1039/B708995C} {\bibfield  {journal}
  {\bibinfo  {journal} {Phys. Chem. Chem. Phys.}\ }\textbf {\bibinfo {volume}
  {9}},\ \bibinfo {pages} {5067} (\bibinfo {year} {2007})}\BibitemShut
  {NoStop}%
\bibitem [{\citenamefont {J\"{u}licher}\ \emph {et~al.}(1997)\citenamefont
  {J\"{u}licher}, \citenamefont {Ajdari},\ and\ \citenamefont
  {Prost}}]{JulAjdPro97}%
  \BibitemOpen
  \bibfield  {author} {\bibinfo {author} {\bibfnamefont {F.}~\bibnamefont
  {J\"{u}licher}}, \bibinfo {author} {\bibfnamefont {A.}~\bibnamefont
  {Ajdari}}, \ and\ \bibinfo {author} {\bibfnamefont {J.}~\bibnamefont
  {Prost}},\ }\href {\doibase 10.1103/RevModPhys.69.1269} {\bibfield  {journal}
  {\bibinfo  {journal} {Rev. Mod. Phys.}\ }\textbf {\bibinfo {volume} {69}},\
  \bibinfo {pages} {1269} (\bibinfo {year} {1997})}\BibitemShut {NoStop}%
\bibitem [{\citenamefont {Bermudez}\ \emph {et~al.}(2000)\citenamefont
  {Bermudez}, \citenamefont {Capron}, \citenamefont {Gase}, \citenamefont
  {Gatti}, \citenamefont {Kajzar}, \citenamefont {Leigh}, \citenamefont
  {Zerbetto},\ and\ \citenamefont {Zhang}}]{Bermudez2000}%
  \BibitemOpen
  \bibfield  {author} {\bibinfo {author} {\bibfnamefont {V.}~\bibnamefont
  {Bermudez}}, \bibinfo {author} {\bibfnamefont {N.}~\bibnamefont {Capron}},
  \bibinfo {author} {\bibfnamefont {T.}~\bibnamefont {Gase}}, \bibinfo {author}
  {\bibfnamefont {F.~G.}\ \bibnamefont {Gatti}}, \bibinfo {author}
  {\bibfnamefont {F.}~\bibnamefont {Kajzar}}, \bibinfo {author} {\bibfnamefont
  {D.~A.}\ \bibnamefont {Leigh}}, \bibinfo {author} {\bibfnamefont
  {F.}~\bibnamefont {Zerbetto}}, \ and\ \bibinfo {author} {\bibfnamefont
  {S.}~\bibnamefont {Zhang}},\ }\href {\doibase 10.1038/35020531} {\bibfield
  {journal} {\bibinfo  {journal} {Nature}\ }\textbf {\bibinfo {volume} {406}},\
  \bibinfo {pages} {608} (\bibinfo {year} {2000})}\BibitemShut {NoStop}%
\bibitem [{\citenamefont {Villegas}\ \emph {et~al.}(2003)\citenamefont
  {Villegas}, \citenamefont {Savel'ev}, \citenamefont {Nori}, \citenamefont
  {Gonzalez}, \citenamefont {Anguita}, \citenamefont {García},\ and\
  \citenamefont {Vicent}}]{Villegas2003}%
  \BibitemOpen
  \bibfield  {author} {\bibinfo {author} {\bibfnamefont {J.~E.}\ \bibnamefont
  {Villegas}}, \bibinfo {author} {\bibfnamefont {S.}~\bibnamefont {Savel'ev}},
  \bibinfo {author} {\bibfnamefont {F.}~\bibnamefont {Nori}}, \bibinfo {author}
  {\bibfnamefont {E.~M.}\ \bibnamefont {Gonzalez}}, \bibinfo {author}
  {\bibfnamefont {J.~V.}\ \bibnamefont {Anguita}}, \bibinfo {author}
  {\bibfnamefont {R.}~\bibnamefont {García}}, \ and\ \bibinfo {author}
  {\bibfnamefont {J.~L.}\ \bibnamefont {Vicent}},\ }\href {\doibase
  10.1126/science.1090390} {\bibfield  {journal} {\bibinfo  {journal}
  {Science}\ }\textbf {\bibinfo {volume} {302}},\ \bibinfo {pages} {1188}
  (\bibinfo {year} {2003})}\BibitemShut {NoStop}%
\bibitem [{\citenamefont {H\"anggi}\ \emph {et~al.}(2005)\citenamefont
  {H\"anggi}, \citenamefont {Marchesoni},\ and\ \citenamefont
  {Nori}}]{Hanggi_2005}%
  \BibitemOpen
  \bibfield  {author} {\bibinfo {author} {\bibfnamefont {P.}~\bibnamefont
  {H\"anggi}}, \bibinfo {author} {\bibfnamefont {F.}~\bibnamefont
  {Marchesoni}}, \ and\ \bibinfo {author} {\bibfnamefont {F.}~\bibnamefont
  {Nori}},\ }\href {\doibase https://doi.org/10.1002/andp.200551701-304}
  {\bibfield  {journal} {\bibinfo  {journal} {Ann. Phys. (Berl.)}\ }\textbf
  {\bibinfo {volume} {517}},\ \bibinfo {pages} {51} (\bibinfo {year}
  {2005})}\BibitemShut {NoStop}%
\bibitem [{\citenamefont {Sj\"olund}\ \emph {et~al.}(2006)\citenamefont
  {Sj\"olund}, \citenamefont {Petra}, \citenamefont {Dion}, \citenamefont
  {Jonsell}, \citenamefont {Nyl\'en}, \citenamefont {Sanchez-Palencia},\ and\
  \citenamefont {Kastberg}}]{SjoPetDio2006}%
  \BibitemOpen
  \bibfield  {author} {\bibinfo {author} {\bibfnamefont {P.}~\bibnamefont
  {Sj\"olund}}, \bibinfo {author} {\bibfnamefont {S.~J.~H.}\ \bibnamefont
  {Petra}}, \bibinfo {author} {\bibfnamefont {C.~M.}\ \bibnamefont {Dion}},
  \bibinfo {author} {\bibfnamefont {S.}~\bibnamefont {Jonsell}}, \bibinfo
  {author} {\bibfnamefont {M.}~\bibnamefont {Nyl\'en}}, \bibinfo {author}
  {\bibfnamefont {L.}~\bibnamefont {Sanchez-Palencia}}, \ and\ \bibinfo
  {author} {\bibfnamefont {A.}~\bibnamefont {Kastberg}},\ }\href {\doibase
  10.1103/PhysRevLett.96.190602} {\bibfield  {journal} {\bibinfo  {journal}
  {Phys. Rev. Lett.}\ }\textbf {\bibinfo {volume} {96}},\ \bibinfo {pages}
  {190602} (\bibinfo {year} {2006})}\BibitemShut {NoStop}%
\bibitem [{\citenamefont {Serreli}\ \emph {et~al.}(2007)\citenamefont
  {Serreli}, \citenamefont {Lee}, \citenamefont {Kay},\ and\ \citenamefont
  {Leigh}}]{Serreli2007}%
  \BibitemOpen
  \bibfield  {author} {\bibinfo {author} {\bibfnamefont {V.}~\bibnamefont
  {Serreli}}, \bibinfo {author} {\bibfnamefont {C.-F.}\ \bibnamefont {Lee}},
  \bibinfo {author} {\bibfnamefont {E.~R.}\ \bibnamefont {Kay}}, \ and\
  \bibinfo {author} {\bibfnamefont {D.~A.}\ \bibnamefont {Leigh}},\ }\href
  {\doibase 10.1038/nature05452} {\bibfield  {journal} {\bibinfo  {journal}
  {Nature}\ }\textbf {\bibinfo {volume} {445}},\ \bibinfo {pages} {523}
  (\bibinfo {year} {2007})}\BibitemShut {NoStop}%
\bibitem [{\citenamefont {Mahmud}\ \emph {et~al.}(2009)\citenamefont {Mahmud},
  \citenamefont {Campbell}, \citenamefont {Bishop}, \citenamefont {Komarova},
  \citenamefont {Chaga}, \citenamefont {Soh}, \citenamefont {Huda},
  \citenamefont {Kandere-Grzybowska},\ and\ \citenamefont
  {Grzybowski}}]{Mahmud2009}%
  \BibitemOpen
  \bibfield  {author} {\bibinfo {author} {\bibfnamefont {G.}~\bibnamefont
  {Mahmud}}, \bibinfo {author} {\bibfnamefont {C.~J.}\ \bibnamefont
  {Campbell}}, \bibinfo {author} {\bibfnamefont {K.~J.~M.}\ \bibnamefont
  {Bishop}}, \bibinfo {author} {\bibfnamefont {Y.~A.}\ \bibnamefont
  {Komarova}}, \bibinfo {author} {\bibfnamefont {O.}~\bibnamefont {Chaga}},
  \bibinfo {author} {\bibfnamefont {S.}~\bibnamefont {Soh}}, \bibinfo {author}
  {\bibfnamefont {S.}~\bibnamefont {Huda}}, \bibinfo {author} {\bibfnamefont
  {K.}~\bibnamefont {Kandere-Grzybowska}}, \ and\ \bibinfo {author}
  {\bibfnamefont {B.~A.}\ \bibnamefont {Grzybowski}},\ }\href {\doibase
  10.1038/nphys1306} {\bibfield  {journal} {\bibinfo  {journal} {Nat. Phys.}\
  }\textbf {\bibinfo {volume} {5}},\ \bibinfo {pages} {606} (\bibinfo {year}
  {2009})}\BibitemShut {NoStop}%
\bibitem [{\citenamefont {Cubero}\ and\ \citenamefont
  {Renzoni}(2016)}]{CubRen2016}%
  \BibitemOpen
  \bibfield  {author} {\bibinfo {author} {\bibfnamefont {D.}~\bibnamefont
  {Cubero}}\ and\ \bibinfo {author} {\bibfnamefont {F.}~\bibnamefont
  {Renzoni}},\ }\href@noop {} {\emph {\bibinfo {title} {Brownian Ratchets: From
  Statistical Physics to Bio and Nano-motors}}}\ (\bibinfo  {publisher}
  {Cambridge University Press},\ \bibinfo {year} {2016})\BibitemShut {NoStop}%
\bibitem [{\citenamefont {Arzola}\ \emph {et~al.}(2017)\citenamefont {Arzola},
  \citenamefont {Villasante-Barahona}, \citenamefont {Volke-Sep\'ulveda},
  \citenamefont {J\'akl},\ and\ \citenamefont {Zem\'anek}}]{ArzVilVol2017}%
  \BibitemOpen
  \bibfield  {author} {\bibinfo {author} {\bibfnamefont {A.~V.}\ \bibnamefont
  {Arzola}}, \bibinfo {author} {\bibfnamefont {M.}~\bibnamefont
  {Villasante-Barahona}}, \bibinfo {author} {\bibfnamefont {K.}~\bibnamefont
  {Volke-Sep\'ulveda}}, \bibinfo {author} {\bibfnamefont {P.}~\bibnamefont
  {J\'akl}}, \ and\ \bibinfo {author} {\bibfnamefont {P.}~\bibnamefont
  {Zem\'anek}},\ }\href {\doibase 10.1103/PhysRevLett.118.138002} {\bibfield
  {journal} {\bibinfo  {journal} {Phys. Rev. Lett.}\ }\textbf {\bibinfo
  {volume} {118}},\ \bibinfo {pages} {138002} (\bibinfo {year}
  {2017})}\BibitemShut {NoStop}%
\bibitem [{\citenamefont {Mukhopadhyay}\ \emph {et~al.}(2018)\citenamefont
  {Mukhopadhyay}, \citenamefont {Xie}, \citenamefont {Liebchen},\ and\
  \citenamefont {Schmelcher}}]{MukXieLie2018}%
  \BibitemOpen
  \bibfield  {author} {\bibinfo {author} {\bibfnamefont {A.~K.}\ \bibnamefont
  {Mukhopadhyay}}, \bibinfo {author} {\bibfnamefont {T.}~\bibnamefont {Xie}},
  \bibinfo {author} {\bibfnamefont {B.}~\bibnamefont {Liebchen}}, \ and\
  \bibinfo {author} {\bibfnamefont {P.}~\bibnamefont {Schmelcher}},\ }\href
  {\doibase 10.1103/PhysRevE.97.050202} {\bibfield  {journal} {\bibinfo
  {journal} {Phys. Rev. E}\ }\textbf {\bibinfo {volume} {97}},\ \bibinfo
  {pages} {050202} (\bibinfo {year} {2018})}\BibitemShut {NoStop}%
\bibitem [{\citenamefont {Forn{\'e}s}(2021)}]{For2021}%
  \BibitemOpen
  \bibfield  {author} {\bibinfo {author} {\bibfnamefont {J.~A.}\ \bibnamefont
  {Forn{\'e}s}},\ }\href@noop {} {\emph {\bibinfo {title} {Principles of
  Brownian and Molecular Motors}}}\ (\bibinfo  {publisher} {Springer},\
  \bibinfo {year} {2021})\BibitemShut {NoStop}%
\bibitem [{\citenamefont {Carlo}\ \emph {et~al.}(2005)\citenamefont {Carlo},
  \citenamefont {Benenti}, \citenamefont {Casati},\ and\ \citenamefont
  {Shepelyansky}}]{Carlo_2005}%
  \BibitemOpen
  \bibfield  {author} {\bibinfo {author} {\bibfnamefont {G.~G.}\ \bibnamefont
  {Carlo}}, \bibinfo {author} {\bibfnamefont {G.}~\bibnamefont {Benenti}},
  \bibinfo {author} {\bibfnamefont {G.}~\bibnamefont {Casati}}, \ and\ \bibinfo
  {author} {\bibfnamefont {D.~L.}\ \bibnamefont {Shepelyansky}},\ }\href
  {\doibase 10.1103/PhysRevLett.94.164101} {\bibfield  {journal} {\bibinfo
  {journal} {Phys. Rev. Lett.}\ }\textbf {\bibinfo {volume} {94}},\ \bibinfo
  {pages} {164101} (\bibinfo {year} {2005})}\BibitemShut {NoStop}%
\bibitem [{\citenamefont {Linke}\ \emph {et~al.}(1999)\citenamefont {Linke},
  \citenamefont {Humphrey}, \citenamefont {Löfgren}, \citenamefont {Sushkov},
  \citenamefont {Newbury}, \citenamefont {Taylor},\ and\ \citenamefont
  {Omling}}]{Linke_1999}%
  \BibitemOpen
  \bibfield  {author} {\bibinfo {author} {\bibfnamefont {H.}~\bibnamefont
  {Linke}}, \bibinfo {author} {\bibfnamefont {T.~E.}\ \bibnamefont {Humphrey}},
  \bibinfo {author} {\bibfnamefont {A.}~\bibnamefont {Löfgren}}, \bibinfo
  {author} {\bibfnamefont {A.~O.}\ \bibnamefont {Sushkov}}, \bibinfo {author}
  {\bibfnamefont {R.}~\bibnamefont {Newbury}}, \bibinfo {author} {\bibfnamefont
  {R.~P.}\ \bibnamefont {Taylor}}, \ and\ \bibinfo {author} {\bibfnamefont
  {P.}~\bibnamefont {Omling}},\ }\href {\doibase 10.1126/science.286.5448.2314}
  {\bibfield  {journal} {\bibinfo  {journal} {Science}\ }\textbf {\bibinfo
  {volume} {286}},\ \bibinfo {pages} {2314} (\bibinfo {year}
  {1999})}\BibitemShut {NoStop}%
\bibitem [{\citenamefont {Lau}\ and\ \citenamefont {Kedem}(2020)}]{LauKed2020}%
  \BibitemOpen
  \bibfield  {author} {\bibinfo {author} {\bibfnamefont {B.}~\bibnamefont
  {Lau}}\ and\ \bibinfo {author} {\bibfnamefont {O.}~\bibnamefont {Kedem}},\
  }\href {\doibase 10.1063/5.0009561} {\bibfield  {journal} {\bibinfo
  {journal} {J. Chem. Phys.}\ }\textbf {\bibinfo {volume} {152}},\ \bibinfo
  {pages} {200901} (\bibinfo {year} {2020})}\BibitemShut {NoStop}%
\bibitem [{\citenamefont {Yoshida}\ \emph {et~al.}(2012)\citenamefont
  {Yoshida}, \citenamefont {Ekins-Daukes}, \citenamefont {Farrell},\ and\
  \citenamefont {Phillips}}]{YosEkiFar2012}%
  \BibitemOpen
  \bibfield  {author} {\bibinfo {author} {\bibfnamefont {M.}~\bibnamefont
  {Yoshida}}, \bibinfo {author} {\bibfnamefont {N.~J.}\ \bibnamefont
  {Ekins-Daukes}}, \bibinfo {author} {\bibfnamefont {D.~J.}\ \bibnamefont
  {Farrell}}, \ and\ \bibinfo {author} {\bibfnamefont {C.~C.}\ \bibnamefont
  {Phillips}},\ }\href {\doibase 10.1063/1.4731277} {\bibfield  {journal}
  {\bibinfo  {journal} {Appl. Phys. Lett.}\ }\textbf {\bibinfo {volume}
  {100}},\ \bibinfo {pages} {263902} (\bibinfo {year} {2012})}\BibitemShut
  {NoStop}%
\bibitem [{\citenamefont {Pusch}\ \emph {et~al.}(2016)\citenamefont {Pusch},
  \citenamefont {Yoshida}, \citenamefont {Hylton}, \citenamefont {Mellor},
  \citenamefont {Phillips}, \citenamefont {Hess},\ and\ \citenamefont
  {Ekins-Daukes}}]{PusYosHyl2016}%
  \BibitemOpen
  \bibfield  {author} {\bibinfo {author} {\bibfnamefont {A.}~\bibnamefont
  {Pusch}}, \bibinfo {author} {\bibfnamefont {M.}~\bibnamefont {Yoshida}},
  \bibinfo {author} {\bibfnamefont {N.~P.}\ \bibnamefont {Hylton}}, \bibinfo
  {author} {\bibfnamefont {A.}~\bibnamefont {Mellor}}, \bibinfo {author}
  {\bibfnamefont {C.~C.}\ \bibnamefont {Phillips}}, \bibinfo {author}
  {\bibfnamefont {O.}~\bibnamefont {Hess}}, \ and\ \bibinfo {author}
  {\bibfnamefont {N.~J.}\ \bibnamefont {Ekins-Daukes}},\ }\href {\doibase
  https://doi.org/10.1002/pip.2751} {\bibfield  {journal} {\bibinfo  {journal}
  {Prog Photovolt}\ }\textbf {\bibinfo {volume} {24}},\ \bibinfo {pages} {656}
  (\bibinfo {year} {2016})}\BibitemShut {NoStop}%
\bibitem [{\citenamefont {Sogabe}\ \emph {et~al.}(2021)\citenamefont {Sogabe},
  \citenamefont {Hung}, \citenamefont {Tamaki}, \citenamefont {Tomi{\'c}},
  \citenamefont {Yamaguchi}, \citenamefont {Ekins-Daukes},\ and\ \citenamefont
  {Okada}}]{SogHunTam2021}%
  \BibitemOpen
  \bibfield  {author} {\bibinfo {author} {\bibfnamefont {T.}~\bibnamefont
  {Sogabe}}, \bibinfo {author} {\bibfnamefont {C.-Y.}\ \bibnamefont {Hung}},
  \bibinfo {author} {\bibfnamefont {R.}~\bibnamefont {Tamaki}}, \bibinfo
  {author} {\bibfnamefont {S.}~\bibnamefont {Tomi{\'c}}}, \bibinfo {author}
  {\bibfnamefont {K.}~\bibnamefont {Yamaguchi}}, \bibinfo {author}
  {\bibfnamefont {N.}~\bibnamefont {Ekins-Daukes}}, \ and\ \bibinfo {author}
  {\bibfnamefont {Y.}~\bibnamefont {Okada}},\ }\href@noop {} {\bibfield
  {journal} {\bibinfo  {journal} {Communications Physics}\ }\textbf {\bibinfo
  {volume} {4}},\ \bibinfo {pages} {1} (\bibinfo {year} {2021})}\BibitemShut
  {NoStop}%
\bibitem [{\citenamefont {Denisov}\ \emph {et~al.}(2007)\citenamefont
  {Denisov}, \citenamefont {Morales-Molina}, \citenamefont {Flach},\ and\
  \citenamefont {H\"anggi}}]{DenMorFla07}%
  \BibitemOpen
  \bibfield  {author} {\bibinfo {author} {\bibfnamefont {S.}~\bibnamefont
  {Denisov}}, \bibinfo {author} {\bibfnamefont {L.}~\bibnamefont
  {Morales-Molina}}, \bibinfo {author} {\bibfnamefont {S.}~\bibnamefont
  {Flach}}, \ and\ \bibinfo {author} {\bibfnamefont {P.}~\bibnamefont
  {H\"anggi}},\ }\href {\doibase 10.1103/PhysRevA.75.063424} {\bibfield
  {journal} {\bibinfo  {journal} {Phys. Rev. A}\ }\textbf {\bibinfo {volume}
  {75}},\ \bibinfo {pages} {063424} (\bibinfo {year} {2007})}\BibitemShut
  {NoStop}%
\bibitem [{\citenamefont {Flach}\ \emph {et~al.}(2000)\citenamefont {Flach},
  \citenamefont {Yevtushenko},\ and\ \citenamefont {Zolotaryuk}}]{FlaYevZol00}%
  \BibitemOpen
  \bibfield  {author} {\bibinfo {author} {\bibfnamefont {S.}~\bibnamefont
  {Flach}}, \bibinfo {author} {\bibfnamefont {O.}~\bibnamefont {Yevtushenko}},
  \ and\ \bibinfo {author} {\bibfnamefont {Y.}~\bibnamefont {Zolotaryuk}},\
  }\href {\doibase 10.1103/PhysRevLett.84.2358} {\bibfield  {journal} {\bibinfo
   {journal} {Phys. Rev. Lett.}\ }\textbf {\bibinfo {volume} {84}},\ \bibinfo
  {pages} {2358} (\bibinfo {year} {2000})}\BibitemShut {NoStop}%
\bibitem [{\citenamefont {Denisov}\ \emph {et~al.}(2014)\citenamefont
  {Denisov}, \citenamefont {Flach},\ and\ \citenamefont
  {H\"anggi}}]{DenFlaHan14}%
  \BibitemOpen
  \bibfield  {author} {\bibinfo {author} {\bibfnamefont {S.}~\bibnamefont
  {Denisov}}, \bibinfo {author} {\bibfnamefont {S.}~\bibnamefont {Flach}}, \
  and\ \bibinfo {author} {\bibfnamefont {P.}~\bibnamefont {H\"anggi}},\ }\href
  {\doibase https://doi.org/10.1016/j.physrep.2014.01.003} {\bibfield
  {journal} {\bibinfo  {journal} {Phys. Rep}\ }\textbf {\bibinfo {volume}
  {538}},\ \bibinfo {pages} {77} (\bibinfo {year} {2014})}\BibitemShut
  {NoStop}%
\bibitem [{\citenamefont {Schanz}\ \emph {et~al.}(2001)\citenamefont {Schanz},
  \citenamefont {Otto}, \citenamefont {Ketzmerick},\ and\ \citenamefont
  {Dittrich}}]{SchOttKet01}%
  \BibitemOpen
  \bibfield  {author} {\bibinfo {author} {\bibfnamefont {H.}~\bibnamefont
  {Schanz}}, \bibinfo {author} {\bibfnamefont {M.-F.}\ \bibnamefont {Otto}},
  \bibinfo {author} {\bibfnamefont {R.}~\bibnamefont {Ketzmerick}}, \ and\
  \bibinfo {author} {\bibfnamefont {T.}~\bibnamefont {Dittrich}},\ }\href
  {\doibase 10.1103/PhysRevLett.87.070601} {\bibfield  {journal} {\bibinfo
  {journal} {Phys. Rev. Lett.}\ }\textbf {\bibinfo {volume} {87}},\ \bibinfo
  {pages} {070601} (\bibinfo {year} {2001})}\BibitemShut {NoStop}%
\bibitem [{\citenamefont {Monteiro}\ \emph {et~al.}(2002)\citenamefont
  {Monteiro}, \citenamefont {Dando}, \citenamefont {Hutchings},\ and\
  \citenamefont {Isherwood}}]{MonDanHut02}%
  \BibitemOpen
  \bibfield  {author} {\bibinfo {author} {\bibfnamefont {T.~S.}\ \bibnamefont
  {Monteiro}}, \bibinfo {author} {\bibfnamefont {P.~A.}\ \bibnamefont {Dando}},
  \bibinfo {author} {\bibfnamefont {N.~A.~C.}\ \bibnamefont {Hutchings}}, \
  and\ \bibinfo {author} {\bibfnamefont {M.~R.}\ \bibnamefont {Isherwood}},\
  }\href {\doibase 10.1103/PhysRevLett.89.194102} {\bibfield  {journal}
  {\bibinfo  {journal} {Phys. Rev. Lett.}\ }\textbf {\bibinfo {volume} {89}},\
  \bibinfo {pages} {194102} (\bibinfo {year} {2002})}\BibitemShut {NoStop}%
\bibitem [{\citenamefont {Jones}\ \emph {et~al.}(2007)\citenamefont {Jones},
  \citenamefont {Goonasekera}, \citenamefont {Meacher}, \citenamefont
  {Jonckheere},\ and\ \citenamefont {Monteiro}}]{JonGooMea07}%
  \BibitemOpen
  \bibfield  {author} {\bibinfo {author} {\bibfnamefont {P.~H.}\ \bibnamefont
  {Jones}}, \bibinfo {author} {\bibfnamefont {M.}~\bibnamefont {Goonasekera}},
  \bibinfo {author} {\bibfnamefont {D.~R.}\ \bibnamefont {Meacher}}, \bibinfo
  {author} {\bibfnamefont {T.}~\bibnamefont {Jonckheere}}, \ and\ \bibinfo
  {author} {\bibfnamefont {T.~S.}\ \bibnamefont {Monteiro}},\ }\href {\doibase
  10.1103/PhysRevLett.98.073002} {\bibfield  {journal} {\bibinfo  {journal}
  {Phys. Rev. Lett.}\ }\textbf {\bibinfo {volume} {98}},\ \bibinfo {pages}
  {073002} (\bibinfo {year} {2007})}\BibitemShut {NoStop}%
\bibitem [{\citenamefont {Lundh}\ and\ \citenamefont
  {Wallin}(2005)}]{LunWal05}%
  \BibitemOpen
  \bibfield  {author} {\bibinfo {author} {\bibfnamefont {E.}~\bibnamefont
  {Lundh}}\ and\ \bibinfo {author} {\bibfnamefont {M.}~\bibnamefont {Wallin}},\
  }\href {\doibase 10.1103/PhysRevLett.94.110603} {\bibfield  {journal}
  {\bibinfo  {journal} {Phys. Rev. Lett.}\ }\textbf {\bibinfo {volume} {94}},\
  \bibinfo {pages} {110603} (\bibinfo {year} {2005})}\BibitemShut {NoStop}%
\bibitem [{\citenamefont {Poletti}\ \emph
  {et~al.}(2007{\natexlab{a}})\citenamefont {Poletti}, \citenamefont {Carlo},\
  and\ \citenamefont {Li}}]{PolCarGab07}%
  \BibitemOpen
  \bibfield  {author} {\bibinfo {author} {\bibfnamefont {D.}~\bibnamefont
  {Poletti}}, \bibinfo {author} {\bibfnamefont {G.~G.}\ \bibnamefont {Carlo}},
  \ and\ \bibinfo {author} {\bibfnamefont {B.}~\bibnamefont {Li}},\ }\href
  {\doibase 10.1103/PhysRevE.75.011102} {\bibfield  {journal} {\bibinfo
  {journal} {Phys. Rev. E}\ }\textbf {\bibinfo {volume} {75}},\ \bibinfo
  {pages} {011102} (\bibinfo {year} {2007}{\natexlab{a}})}\BibitemShut
  {NoStop}%
\bibitem [{\citenamefont {Dana}\ \emph {et~al.}(2008)\citenamefont {Dana},
  \citenamefont {Ramareddy}, \citenamefont {Talukdar},\ and\ \citenamefont
  {Summy}}]{Dana_2008}%
  \BibitemOpen
  \bibfield  {author} {\bibinfo {author} {\bibfnamefont {I.}~\bibnamefont
  {Dana}}, \bibinfo {author} {\bibfnamefont {V.}~\bibnamefont {Ramareddy}},
  \bibinfo {author} {\bibfnamefont {I.}~\bibnamefont {Talukdar}}, \ and\
  \bibinfo {author} {\bibfnamefont {G.~S.}\ \bibnamefont {Summy}},\ }\href
  {\doibase 10.1103/PhysRevLett.100.024103} {\bibfield  {journal} {\bibinfo
  {journal} {Phys. Rev. Lett.}\ }\textbf {\bibinfo {volume} {100}},\ \bibinfo
  {pages} {024103} (\bibinfo {year} {2008})}\BibitemShut {NoStop}%
\bibitem [{\citenamefont {Gong}\ and\ \citenamefont {Brumer}(2004)}]{GonBru04}%
  \BibitemOpen
  \bibfield  {author} {\bibinfo {author} {\bibfnamefont {J.}~\bibnamefont
  {Gong}}\ and\ \bibinfo {author} {\bibfnamefont {P.}~\bibnamefont {Brumer}},\
  }\href {\doibase 10.1103/PhysRevE.70.016202} {\bibfield  {journal} {\bibinfo
  {journal} {Phys. Rev. E}\ }\textbf {\bibinfo {volume} {70}},\ \bibinfo
  {pages} {016202} (\bibinfo {year} {2004})}\BibitemShut {NoStop}%
\bibitem [{\citenamefont {Poletti}\ \emph
  {et~al.}(2007{\natexlab{b}})\citenamefont {Poletti}, \citenamefont {Benenti},
  \citenamefont {Casati},\ and\ \citenamefont {Li}}]{PolBenCas2007}%
  \BibitemOpen
  \bibfield  {author} {\bibinfo {author} {\bibfnamefont {D.}~\bibnamefont
  {Poletti}}, \bibinfo {author} {\bibfnamefont {G.}~\bibnamefont {Benenti}},
  \bibinfo {author} {\bibfnamefont {G.}~\bibnamefont {Casati}}, \ and\ \bibinfo
  {author} {\bibfnamefont {B.}~\bibnamefont {Li}},\ }\href {\doibase
  10.1103/PhysRevA.76.023421} {\bibfield  {journal} {\bibinfo  {journal} {Phys.
  Rev. A}\ }\textbf {\bibinfo {volume} {76}},\ \bibinfo {pages} {023421}
  (\bibinfo {year} {2007}{\natexlab{b}})}\BibitemShut {NoStop}%
\bibitem [{\citenamefont {Zhao}\ \emph {et~al.}(2016)\citenamefont {Zhao},
  \citenamefont {Ding}, \citenamefont {Liu},\ and\ \citenamefont
  {Fu}}]{ZhaDinLiu2016}%
  \BibitemOpen
  \bibfield  {author} {\bibinfo {author} {\bibfnamefont {W.-L.}\ \bibnamefont
  {Zhao}}, \bibinfo {author} {\bibfnamefont {C.-Y.}\ \bibnamefont {Ding}},
  \bibinfo {author} {\bibfnamefont {J.}~\bibnamefont {Liu}}, \ and\ \bibinfo
  {author} {\bibfnamefont {L.-B.}\ \bibnamefont {Fu}},\ }\href {\doibase
  10.1088/0953-4075/49/12/125303} {\bibfield  {journal} {\bibinfo  {journal}
  {J. Phys. B}\ }\textbf {\bibinfo {volume} {49}},\ \bibinfo {pages} {125303}
  (\bibinfo {year} {2016})}\BibitemShut {NoStop}%
\bibitem [{\citenamefont {Salger}\ \emph {et~al.}(2009)\citenamefont {Salger},
  \citenamefont {Kling}, \citenamefont {Hecking}, \citenamefont {Geckeler},
  \citenamefont {Morales-Molina},\ and\ \citenamefont {Weitz}}]{SalKliHec2009}%
  \BibitemOpen
  \bibfield  {author} {\bibinfo {author} {\bibfnamefont {T.}~\bibnamefont
  {Salger}}, \bibinfo {author} {\bibfnamefont {S.}~\bibnamefont {Kling}},
  \bibinfo {author} {\bibfnamefont {T.}~\bibnamefont {Hecking}}, \bibinfo
  {author} {\bibfnamefont {C.}~\bibnamefont {Geckeler}}, \bibinfo {author}
  {\bibfnamefont {L.}~\bibnamefont {Morales-Molina}}, \ and\ \bibinfo {author}
  {\bibfnamefont {M.}~\bibnamefont {Weitz}},\ }\href {\doibase
  10.1126/science.1179546} {\bibfield  {journal} {\bibinfo  {journal}
  {Science}\ }\textbf {\bibinfo {volume} {326}},\ \bibinfo {pages} {1241}
  (\bibinfo {year} {2009})}\BibitemShut {NoStop}%
\bibitem [{\citenamefont {Valdez}\ \emph {et~al.}(2018)\citenamefont {Valdez},
  \citenamefont {Shchedrin}, \citenamefont {Heimsoth}, \citenamefont
  {Creffield}, \citenamefont {Sols},\ and\ \citenamefont
  {Carr}}]{ValShcHel2018}%
  \BibitemOpen
  \bibfield  {author} {\bibinfo {author} {\bibfnamefont {M.~A.}\ \bibnamefont
  {Valdez}}, \bibinfo {author} {\bibfnamefont {G.}~\bibnamefont {Shchedrin}},
  \bibinfo {author} {\bibfnamefont {M.}~\bibnamefont {Heimsoth}}, \bibinfo
  {author} {\bibfnamefont {C.~E.}\ \bibnamefont {Creffield}}, \bibinfo {author}
  {\bibfnamefont {F.}~\bibnamefont {Sols}}, \ and\ \bibinfo {author}
  {\bibfnamefont {L.~D.}\ \bibnamefont {Carr}},\ }\href {\doibase
  10.1103/PhysRevLett.120.234101} {\bibfield  {journal} {\bibinfo  {journal}
  {Phys. Rev. Lett.}\ }\textbf {\bibinfo {volume} {120}},\ \bibinfo {pages}
  {234101} (\bibinfo {year} {2018})}\BibitemShut {NoStop}%
\bibitem [{\citenamefont {Valdez}\ \emph {et~al.}(2019)\citenamefont {Valdez},
  \citenamefont {Shchedrin}, \citenamefont {Sols},\ and\ \citenamefont
  {Carr}}]{ValShcSol2019}%
  \BibitemOpen
  \bibfield  {author} {\bibinfo {author} {\bibfnamefont {M.~A.}\ \bibnamefont
  {Valdez}}, \bibinfo {author} {\bibfnamefont {G.}~\bibnamefont {Shchedrin}},
  \bibinfo {author} {\bibfnamefont {F.}~\bibnamefont {Sols}}, \ and\ \bibinfo
  {author} {\bibfnamefont {L.~D.}\ \bibnamefont {Carr}},\ }\href {\doibase
  10.1103/PhysRevA.99.063609} {\bibfield  {journal} {\bibinfo  {journal} {Phys.
  Rev. A}\ }\textbf {\bibinfo {volume} {99}},\ \bibinfo {pages} {063609}
  (\bibinfo {year} {2019})}\BibitemShut {NoStop}%
\bibitem [{\citenamefont {Paul}\ and\ \citenamefont
  {B\"acker}(2020)}]{SanArn20}%
  \BibitemOpen
  \bibfield  {author} {\bibinfo {author} {\bibfnamefont {S.}~\bibnamefont
  {Paul}}\ and\ \bibinfo {author} {\bibfnamefont {A.}~\bibnamefont
  {B\"acker}},\ }\href {\doibase 10.1103/PhysRevE.102.050102} {\bibfield
  {journal} {\bibinfo  {journal} {Phys. Rev. E}\ }\textbf {\bibinfo {volume}
  {102}},\ \bibinfo {pages} {050102} (\bibinfo {year} {2020})}\BibitemShut
  {NoStop}%
\bibitem [{\citenamefont {Lichtenberg}\ and\ \citenamefont
  {Lieberman}(2013)}]{LicLie2013}%
  \BibitemOpen
  \bibfield  {author} {\bibinfo {author} {\bibfnamefont {A.~J.}\ \bibnamefont
  {Lichtenberg}}\ and\ \bibinfo {author} {\bibfnamefont {M.~A.}\ \bibnamefont
  {Lieberman}},\ }\href@noop {} {\emph {\bibinfo {title} {Regular and
  Stochastic Motion}}},\ Vol.\ \bibinfo {volume} {38 of Applied Mathematical
  Sciences}\ (\bibinfo  {publisher} {Springer Science \& Business Media},\
  \bibinfo {year} {2013})\BibitemShut {NoStop}%
\bibitem [{\citenamefont {Gardiner}\ \emph {et~al.}(1997)\citenamefont
  {Gardiner}, \citenamefont {Cirac},\ and\ \citenamefont
  {Zoller}}]{Gardiner_1997}%
  \BibitemOpen
  \bibfield  {author} {\bibinfo {author} {\bibfnamefont {S.~A.}\ \bibnamefont
  {Gardiner}}, \bibinfo {author} {\bibfnamefont {J.~I.}\ \bibnamefont {Cirac}},
  \ and\ \bibinfo {author} {\bibfnamefont {P.}~\bibnamefont {Zoller}},\ }\href
  {\doibase 10.1103/PhysRevLett.79.4790} {\bibfield  {journal} {\bibinfo
  {journal} {Phys. Rev. Lett.}\ }\textbf {\bibinfo {volume} {79}},\ \bibinfo
  {pages} {4790} (\bibinfo {year} {1997})}\BibitemShut {NoStop}%
\bibitem [{\citenamefont {Sankaranarayanan}\ \emph {et~al.}(2001)\citenamefont
  {Sankaranarayanan}, \citenamefont {Lakshminarayan},\ and\ \citenamefont
  {Sheorey}}]{Sankarnarayanan_2001}%
  \BibitemOpen
  \bibfield  {author} {\bibinfo {author} {\bibfnamefont {R.}~\bibnamefont
  {Sankaranarayanan}}, \bibinfo {author} {\bibfnamefont {A.}~\bibnamefont
  {Lakshminarayan}}, \ and\ \bibinfo {author} {\bibfnamefont {V.~B.}\
  \bibnamefont {Sheorey}},\ }\href {\doibase 10.1103/PhysRevE.64.046210}
  {\bibfield  {journal} {\bibinfo  {journal} {Phys. Rev. E}\ }\textbf {\bibinfo
  {volume} {64}},\ \bibinfo {pages} {046210} (\bibinfo {year}
  {2001})}\BibitemShut {NoStop}%
\bibitem [{\citenamefont {Paul}\ \emph {et~al.}(2016)\citenamefont {Paul},
  \citenamefont {Pal},\ and\ \citenamefont {Santhanam}}]{Sanku_2016}%
  \BibitemOpen
  \bibfield  {author} {\bibinfo {author} {\bibfnamefont {S.}~\bibnamefont
  {Paul}}, \bibinfo {author} {\bibfnamefont {H.}~\bibnamefont {Pal}}, \ and\
  \bibinfo {author} {\bibfnamefont {M.~S.}\ \bibnamefont {Santhanam}},\ }\href
  {\doibase 10.1103/PhysRevE.93.060203} {\bibfield  {journal} {\bibinfo
  {journal} {Phys. Rev. E}\ }\textbf {\bibinfo {volume} {93}},\ \bibinfo
  {pages} {060203} (\bibinfo {year} {2016})}\BibitemShut {NoStop}%
\bibitem [{\citenamefont {Paul}\ and\ \citenamefont
  {Santhanam}(2018)}]{Sanku_2018}%
  \BibitemOpen
  \bibfield  {author} {\bibinfo {author} {\bibfnamefont {S.}~\bibnamefont
  {Paul}}\ and\ \bibinfo {author} {\bibfnamefont {M.~S.}\ \bibnamefont
  {Santhanam}},\ }\href {\doibase 10.1103/PhysRevE.97.032217} {\bibfield
  {journal} {\bibinfo  {journal} {Phys. Rev. E}\ }\textbf {\bibinfo {volume}
  {97}},\ \bibinfo {pages} {032217} (\bibinfo {year} {2018})}\BibitemShut
  {NoStop}%
\bibitem [{\citenamefont {Casati}\ \emph {et~al.}(1979)\citenamefont {Casati},
  \citenamefont {Chirikov}, \citenamefont {Izraelev},\ and\ \citenamefont
  {Ford}}]{Casati_1979}%
  \BibitemOpen
  \bibfield  {author} {\bibinfo {author} {\bibfnamefont {G.}~\bibnamefont
  {Casati}}, \bibinfo {author} {\bibfnamefont {B.~V.}\ \bibnamefont
  {Chirikov}}, \bibinfo {author} {\bibfnamefont {F.~M.}\ \bibnamefont
  {Izraelev}}, \ and\ \bibinfo {author} {\bibfnamefont {J.}~\bibnamefont
  {Ford}},\ }in\ \href {https://link.springer.com/chapter/10.1007/BFb0021757}
  {\emph {\bibinfo {booktitle} {Stochastic Behavior in Classical and Quantum
  Hamiltonian Systems}}},\ \bibinfo {editor} {edited by\ \bibinfo {editor}
  {\bibfnamefont {G.}~\bibnamefont {Casati}}\ and\ \bibinfo {editor}
  {\bibfnamefont {J.}~\bibnamefont {Ford}}}\ (\bibinfo  {publisher} {Springer
  Berlin Heidelberg},\ \bibinfo {address} {Berlin, Heidelberg},\ \bibinfo
  {year} {1979})\ pp.\ \bibinfo {pages} {334--352}\BibitemShut {NoStop}%
\bibitem [{\citenamefont {Izrailev}(1990)}]{IZRAILEV1990}%
  \BibitemOpen
  \bibfield  {author} {\bibinfo {author} {\bibfnamefont {F.~M.}\ \bibnamefont
  {Izrailev}},\ }\href {\doibase https://doi.org/10.1016/0370-1573(90)90067-C}
  {\bibfield  {journal} {\bibinfo  {journal} {Phys. Rep}\ }\textbf {\bibinfo
  {volume} {196}},\ \bibinfo {pages} {299} (\bibinfo {year}
  {1990})}\BibitemShut {NoStop}%
\bibitem [{\citenamefont {Santhanam}\ \emph {et~al.}(2022)\citenamefont
  {Santhanam}, \citenamefont {Paul},\ and\ \citenamefont
  {Kannan}}]{SanPauKan2022}%
  \BibitemOpen
  \bibfield  {author} {\bibinfo {author} {\bibfnamefont {M.~S.}\ \bibnamefont
  {Santhanam}}, \bibinfo {author} {\bibfnamefont {S.}~\bibnamefont {Paul}}, \
  and\ \bibinfo {author} {\bibfnamefont {J.~B.}\ \bibnamefont {Kannan}},\
  }\href {\doibase https://doi.org/10.1016/j.physrep.2022.01.002} {\bibfield
  {journal} {\bibinfo  {journal} {Phys. Rep}\ }\textbf {\bibinfo {volume}
  {956}},\ \bibinfo {pages} {1} (\bibinfo {year} {2022})}\BibitemShut {NoStop}%
\bibitem [{\citenamefont {Fishman}\ \emph {et~al.}(1982)\citenamefont
  {Fishman}, \citenamefont {Grempel},\ and\ \citenamefont
  {Prange}}]{FisShmGre82}%
  \BibitemOpen
  \bibfield  {author} {\bibinfo {author} {\bibfnamefont {S.}~\bibnamefont
  {Fishman}}, \bibinfo {author} {\bibfnamefont {D.~R.}\ \bibnamefont
  {Grempel}}, \ and\ \bibinfo {author} {\bibfnamefont {R.~E.}\ \bibnamefont
  {Prange}},\ }\href {\doibase 10.1103/PhysRevLett.49.509} {\bibfield
  {journal} {\bibinfo  {journal} {Phys. Rev. Lett.}\ }\textbf {\bibinfo
  {volume} {49}},\ \bibinfo {pages} {509} (\bibinfo {year} {1982})}\BibitemShut
  {NoStop}%
\bibitem [{\citenamefont {Sokoloff}(1985)}]{Sokoloff_1985}%
  \BibitemOpen
  \bibfield  {author} {\bibinfo {author} {\bibfnamefont {J.}~\bibnamefont
  {Sokoloff}},\ }\href {\doibase https://doi.org/10.1016/0370-1573(85)90088-2}
  {\bibfield  {journal} {\bibinfo  {journal} {Phys. Rep}\ }\textbf {\bibinfo
  {volume} {126}},\ \bibinfo {pages} {189} (\bibinfo {year}
  {1985})}\BibitemShut {NoStop}%
\bibitem [{\citenamefont {Lima}\ and\ \citenamefont
  {Shepelyansky}(1991)}]{Lima_1991}%
  \BibitemOpen
  \bibfield  {author} {\bibinfo {author} {\bibfnamefont {R.}~\bibnamefont
  {Lima}}\ and\ \bibinfo {author} {\bibfnamefont {D.}~\bibnamefont
  {Shepelyansky}},\ }\href {\doibase 10.1103/PhysRevLett.67.1377} {\bibfield
  {journal} {\bibinfo  {journal} {Phys. Rev. Lett.}\ }\textbf {\bibinfo
  {volume} {67}},\ \bibinfo {pages} {1377} (\bibinfo {year}
  {1991})}\BibitemShut {NoStop}%
\bibitem [{\citenamefont {Lakshminarayan}(2001)}]{Arul_2001}%
  \BibitemOpen
  \bibfield  {author} {\bibinfo {author} {\bibfnamefont {A.}~\bibnamefont
  {Lakshminarayan}},\ }\href {\doibase 10.1103/PhysRevE.64.036207} {\bibfield
  {journal} {\bibinfo  {journal} {Phys. Rev. E}\ }\textbf {\bibinfo {volume}
  {64}},\ \bibinfo {pages} {036207} (\bibinfo {year} {2001})}\BibitemShut
  {NoStop}%
\bibitem [{\citenamefont {Gadway}\ \emph {et~al.}(2013)\citenamefont {Gadway},
  \citenamefont {Reeves}, \citenamefont {Krinner},\ and\ \citenamefont
  {Schneble}}]{Gadway_2013}%
  \BibitemOpen
  \bibfield  {author} {\bibinfo {author} {\bibfnamefont {B.}~\bibnamefont
  {Gadway}}, \bibinfo {author} {\bibfnamefont {J.}~\bibnamefont {Reeves}},
  \bibinfo {author} {\bibfnamefont {L.}~\bibnamefont {Krinner}}, \ and\
  \bibinfo {author} {\bibfnamefont {D.}~\bibnamefont {Schneble}},\ }\href
  {\doibase 10.1103/PhysRevLett.110.190401} {\bibfield  {journal} {\bibinfo
  {journal} {Phys. Rev. Lett.}\ }\textbf {\bibinfo {volume} {110}},\ \bibinfo
  {pages} {190401} (\bibinfo {year} {2013})}\BibitemShut {NoStop}%
\end{thebibliography}%

	

\section*{Supplementary material}

# Supplementary material
# Interaction-induced directed transport in quantum chaotic systems

Sanku Paul,[1] J. Bharathi Kannan,[2] and M. S. Santhanam[2]

[1]*Department of Physics and Astronomy, Michigan State University, East Lansing, Michigan 48824, USA*
[2]*Indian Institute of Science Education and Research, Dr. Homi Bhabha Road, Pune 411 008, India*

This Supplemental Material provides additional details for the results stated in the main text. In Sec. I, the interaction-induced time-reversal symmetry breaking in subsystems is discussed. Section II provides details of the stroboscopic map of the coupled kicked rotor in Eq. (2), and the coupled kicked Harper model in Eq. (3) of the main paper. The behaviour of the classical current in correspondence to that of the quantum current shown in Fig. 1 of the main paper is discussed in Sec. III.

## I. TIME-REVERSAL SYMMETRY BREAKING

Let us consider a coupled system consisting of two sub-systems which are interacting with one another. In this section, we analytically show the effect of interaction in breaking the time-reversal symmetry of a subsystem in the presence of the other. The formalism introduced is general, not limited to the models considered in the main paper, and can be extended to many-body systems as well. To begin with, let us first understand the situation in the absence of interaction and then discuss the role of interaction in time-reversal symmetry breaking. The subsequent subsections discuss both situations.

### A. Without interaction

Here, we discuss the time-reversal symmetry of the individual subsystems in the absence of interactions. To this end, we consider the initial state $|\psi(0)\rangle$ of the interacting system at time $t = 0$ to be a product state of the subsystems

$$|\psi(0)\rangle = |\phi_1(0)\rangle \otimes |\phi_2(0)\rangle = \sum_i a_i |i\rangle_1 \otimes \sum_j b_j |j\rangle_2 \,, \tag{S.1}$$

where $|i\rangle_1$ ($|j\rangle_2$) represents the basis of subsystem-1 (subsystem-2). The coefficients $a_i$ ($b_j$) satisfies the normalization condition $\sum_i |a_i|^2 = 1$ ($\sum_j |b_j|^2 = 1$).

In the case of an uncoupled system, the time evolution operator of the entire system $\mathcal{U}$ can be written as the tensor product of the individual unitary time evolution operators $U_1$ and $U_2$ at all times

$$\mathcal{U} = U_1 \otimes U_2 \,. \tag{S.2}$$

The time evolved state of the composite system at time $\tau > 0$ is given by

$$|\psi(\tau)\rangle = \mathcal{U}^\tau |\psi(0)\rangle = U_1^\tau |\phi_1(0)\rangle \otimes U_2^\tau |\phi_2(0)\rangle \,. \tag{S.3}$$

Now to examine the time-reversal symmetry of the subsystem in the presence of the other, we evolve subsystem-1 backward in time for the duration $-\tau$, while the subsystem-2 is evolved forward for duration $\tau$, the state obtained is

$$\begin{aligned} |\psi(\tau')\rangle &= (U_1^{-\tau} \otimes U_2^\tau) |\psi(\tau)\rangle \\ &= (U_1^{-\tau} \otimes U_2^\tau) U_1^\tau |\phi_1(0)\rangle \otimes U_2^\tau |\phi_2(0)\rangle \\ &= U_1^{-\tau} U_1^\tau |\phi_1(0)\rangle \otimes U_2^\tau U_2^\tau |\phi_2(0)\rangle \\ &= |\phi_1(0)\rangle \otimes |\phi_2(2\tau)\rangle \,. \end{aligned} \tag{S.4}$$

The density matrix of the composite system at $\tau'$ can be estimated using the time evolved state obtained in Eq. (S.4) ( where $\tau'$ indicates successive time duration of $(\tau, -\tau)$ for subsystem-1, and $(\tau, \tau)$ for subsystem-2) and is expressed as

$$\rho(\tau') = |\psi(\tau')\rangle\langle\psi(\tau')| \;\; = |\phi_1(0)\rangle\langle\phi_1(0)| \otimes |\phi_2(2\tau)\rangle\langle\phi_2(2\tau)| \,. \tag{S.5}$$

The reduced density matrix of the subsystem-1 can be obtained from $\rho(\tau')$ by tracing out the subsystem-2 at $\tau'$ and is given by

$$\begin{aligned}\rho_1(\tau') &= \text{Tr}_2[\rho(\tau')] \\ &= \sum_j {}_2\langle j|\phi_1(0)\rangle\langle\phi_1(0)| \otimes |\phi_2(2\tau)\rangle\langle\phi_2(2\tau)|j\rangle_2 \\ &= |\phi_1(0)\rangle\langle\phi_1(0)| \otimes \sum_j {}_2\langle j| \sum_{j',j''} b_{j'}(2\tau)b^*_{j''}(2\tau)|j'\rangle_2 \, {}_2\langle j''|j\rangle_2 \\ &= |\phi_1(0)\rangle\langle\phi_1(0)| \sum_j |b_j(2\tau)|^2 \\ &= |\phi_1(0)\rangle\langle\phi_1(0)|, \end{aligned} \quad (S.6)$$

where $b_j = \langle j|\phi_2(2\tau)\rangle$ and $\sum_j |b_j(2\tau)|^2 = 1$. Now to investigate the time reversal symmetry, we calculate the expectation value of $\rho_1(\tau')$ with respect to the initial state of the subsystem-1. In this paper, we call this quantity $Q_\tau(n) = \langle \phi_1(0)|\rho_1(n)|\phi_1(0)\rangle$. Evaluating $Q_\tau(n = \tau')$ leads to

$$\langle\phi_1(0)|\rho_1(\tau')|\phi_1(0)\rangle = \langle\phi_1(0)|\phi_1(0)\rangle\langle\phi_1(0)|\phi_1(0)\rangle = 1. \quad (S.7)$$

$Q_\tau(n = \tau') = 1$ implies that subsystem-1 retraces its path to the initial state after time $\tau'$. Thus, we see that the time-reversal symmetry is preserved in subsystem-1 in the presence of subsystem-2. This is an expected result as there is no interaction between the subsystems and is also a property of unitary evolution. Using a similar formalism discussed above, it can be shown that time-reversal symmetry is preserved in subsystem-2 in the presence of subsystem-1.

### B. With interaction

In this subsection, we show that the interaction between two subsystems breaks the time-reversal symmetry of the individual subsystems. For that, consider the initial state at time $t = 0$ of the interacting system to be a product state,

$$|\psi(0)\rangle = |\phi_1(0)\rangle \otimes |\phi_2(0)\rangle = \sum_i a_i |i\rangle_1 \otimes \sum_j b_j |j\rangle_2 = \sum_{i,j} c_{ij}(0)|i,j\rangle, \quad (S.8)$$

where $a_i, |i\rangle_1, b_j$, and $|j\rangle_2$ have the same meaning as in Eq. (S.1).

The time evolution operator for interacting system of two subsystems can be written as

$$\mathcal{U} = (U_1 \otimes U_2)U_I, \quad (S.9)$$

where $U_1$, $U_2$ are the time evolution operators for subsystem-1 and subsystem-2, respectively, and $U_I$ is the unitary time evolution operator corresponding to the interaction between the subsystems. The time evolved state at $\tau$ is given by

$$\begin{aligned}|\psi(\tau)\rangle &= \mathcal{U}^\tau|\psi(0)\rangle = [(U_1 \otimes U_2)U_I]^\tau \sum_{i,j} c_{ij}(0)|i,j\rangle \\ &= \sum_{i,j} d_{ij}(\tau)|i,j\rangle. \end{aligned} \quad (S.10)$$

Now, to investigate if the time-reversal symmetry of the subsystem-1 in the presence of subsystem-2 is broken or not, we evolve subsystem-1 backward in time, *i.e.* $-\tau$, while subsystem-2 and the interaction term move forward for the same time $\tau$. The evolved state thus obtained is

$$\begin{aligned}|\psi(\tau')\rangle &= (U_1^{-\tau} \otimes U_2^\tau)U_I^\tau|\psi(\tau)\rangle \\ &= [(U_1^{-\tau} \otimes U_2^\tau)U_I^\tau][(U_1 \otimes U_2)U_I]^\tau \sum_{i,j} c_{ij}(0)|i,j\rangle \\ &= [U_1^0 \otimes U_2^{2\tau}]U_I^{2\tau} \sum_{i,j} c_{ij}(0)|i,j\rangle \\ &= \sum_{i,j} d_{ij}(\tau')|i,j\rangle. \end{aligned} \quad (S.11)$$

The density matrix of the total system at $\tau$ is given by

$$\rho(\tau') = |\psi(\tau')\rangle\langle\psi(\tau')| = \sum_{i,j,i',j'} d_{ij}(\tau')d^*_{i'j'}(\tau')|i,j\rangle\langle i',j'|. \tag{S.12}$$

The reduced density matrix of subsystem-1 obtained by tracing out subsystem-2 from the total density matrix $\rho(\tau')$ in Eq. (S.12) is

$$\begin{aligned}\rho_1(\tau') &= \text{Tr}_2[\rho(\tau')] \\ &= \sum_{j''} {}_2\langle j''| \sum_{i,j,i',j'} d_{ij}(\tau')d^*_{i'j'}(\tau')|i,j\rangle\langle i',j'|j''\rangle_2 \\ &= \sum_{i,i'} {}_1|i\rangle\langle i'|_1 \sum_j d_{ij}(\tau')d^*_{i'j}(\tau').\end{aligned} \tag{S.13}$$

Now to understand if the time reversal symmetry is preserved or not, we evaluate $Q_\tau(\tau') = \langle\phi_1(0)|\rho_1(\tau')|\phi_1(0)\rangle$, which is

$$\begin{aligned}Q_\tau(\tau') &= \sum_{i''} a^*_{i''}(0) \; {}_1\langle i''|\sum_{i,i'}|i\rangle_1 \; {}_1\langle i'|\sum_j d_{ij}(\tau')d^*_{i'j}(\tau') \sum_{i'''} a_{i'''}(0)|i'''\rangle_1 \\ &= \sum_{i'',i,i',i'''} a^*_{i''}(0)a_{i'''}(0) \; {}_1\langle i''|i\rangle_1 \langle i'|i'''\rangle_1 \sum_j d_{ij}(\tau')d^*_{i'j}(\tau') \\ &= \sum_{i,i'} a^*_i(0)a_{i'}(0) \sum_j d_{ij}(\tau')d^*_{i'j}(\tau') \\ &\leq 1.\end{aligned} \tag{S.14}$$

In this last equation, equality holds if there is no interaction between the subsystems. Otherwise, $Q_\tau(n = \tau') < 1$. This implies that in the presence of interaction, subsystem-1 does not retrace its path to the initial state after time $\tau'$. This indicates that the time-reversal symmetry is broken in subsystem-1 due to the interaction between the subsystems. The same argument holds for the time-reversal symmetry breaking in subsystem-2.

## II. CLASSICAL STROBOSCOPIC MAP

The classical dynamics of CKR in Eq. (2) and CKH in Eq. (3) of the main paper can be reduced to a stroboscopic map on a cylinder as given below in Eq. (S.15) and Eq. (S.16). The stroboscopic map can be obtained using the Hamilton equations of motion and for the CKR, it is expressed as

$$\begin{aligned} p_1^{n+1} &= p_1^n + K_1^r \sin(q_1^n) - \varepsilon \frac{\partial V_{\text{int}}(q_1,q_2)}{\partial q_1}\Big|_{q_1^n,q_2^n}, \\ p_2^{n+1} &= p_2^n + K_2^r \sin(q_2^n) - \varepsilon \frac{\partial V_{\text{int}}(q_1,q_2)}{\partial q_2}\Big|_{q_1^n,q_2^n} \\ q_1^{n+1} &= (q_1^n + p_1^{n+1}) \mod 2\pi, \text{ and} \\ q_2^{n+1} &= (q_2^n + p_2^{n+1}) \mod 2\pi,\end{aligned} \tag{S.15}$$

where $q_i^n$ ($p_i^n$) represents the position (momentum) of the $i$-th rotor at $t=n$, where $i=1,2$.

The stroboscopic map of the CKH model is

$$\begin{aligned} p_1^{n+1} &= p_1^n + K_1^h \sin(q_1^n) - \varepsilon \frac{\partial V_{\text{int}}(q_1,q_2)}{\partial q_1}\Big|_{q_1^n,q_2^n}, \\ p_2^{n+1} &= p_2^n + K_2^h \sin(q_2^n) - \varepsilon \frac{\partial V_{\text{int}}(q_1,q_2)}{\partial q_2}\Big|_{q_1^n,q_2^n}, \\ q_1^{n+1} &= (q_1^n - L_1^h \sin p_1^{n+1}) \mod 2\pi, \text{ and} \\ q_2^{n+1} &= (q_2^n - L_2^h \sin p_2^{n+1}) \mod 2\pi,\end{aligned} \tag{S.16}$$

where $q_i^n$ ($p_i^n$) represents the position (momentum) of the $i$-th Harper model at $t=n$, where $i=1,2$.

## III. CLASSICAL CURRENT

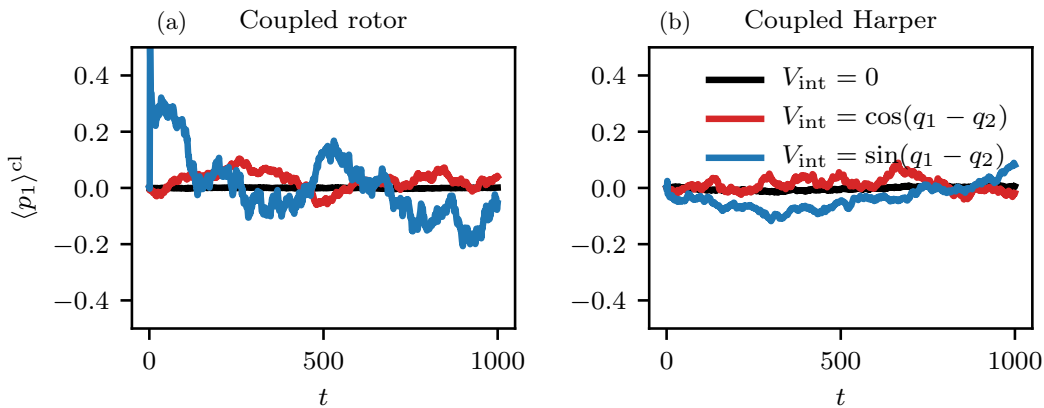

FIG. 1. Absence of averaged classical current $\langle p_1 \rangle^{\text{cl}}$ is shown for (a) CKR (Eq. (2) of the main paper) and (b) CKH (Eq. (3) of the main paper) and for three different interaction potentials with $\varepsilon = 5$: (i) non-interacting ($V_{\text{int}} = 0$) (black cure), (ii) $V_{\text{int}} = \cos(q_1 - q_2)$ (red), and (iii) $V_{\text{int}} = \sin(q_1 - q_2)$ (blue). The averaging is done over $10^6$ initial samples taken from an interval $q_i \in [0, 2\pi]$ and $p_i = 0$. Other parameters are the same as in Fig. 1 of the main paper.

In this section, the behaviour of classical current $\langle p_1 \rangle^{\text{cl}}$ is discussed for CKR in Eq. (2) and CKH in Eq. (3) of the main paper. The classical current for CKR and CKH is numerically simulated using Eq. (S.15) and Eq. (S.16), respectively and is averaged over $10^6$ initial sample points to reduce fluctuations. Figure 1 illustrates the absence of classical current for both the interacting and non-interacting cases. This is in stark contrast to the quantum current displayed in Fig. 1 of the main paper. The absence of the classical current can be attributed to the presence of near-complete chaos in the system and the absence of explicit temporal symmetry breaking.



\end{document}